\documentclass[final,1p,times]{elsarticle}

\usepackage{amssymb}
\usepackage{amsmath}

\journal{Journal of Mechanics and Physics of Solids}

\begin{document}

\begin{frontmatter}



\title{Translation Method Bounds for 3D Isotropic Conductivity--Shear Modulus and the Search for Attaining Microstructures}


\author{Justin Tahmassebpur \corref{Corresponding author}} 
\ead{jt577@cornell.edu}
\affiliation{organization={Department of Applied and Engineering Physics, Cornell University}, city={Ithaca}, postcode={14850}, state={NY}, country={USA}}

\author{Tom\'as Arias} 
\affiliation{organization={Department of Physics, Cornell University}, city={Ithaca}, postcode={14850}, state={NY}, country={USA}}

\begin{abstract}
We compute conductivity--shear-modulus cross-property bounds in three-dimensional isotropic two-phase composites using the translation method, and obtain the tightest bounds reported to date. To assess attainability, we perform unit-cell topology optimization on thousands of distinct initial structures and compare optimized designs to the isotropic bounds using an effective-medium approximation that models the macroscopic material as an isotropic assembly of randomly oriented grains. The best jointly performing designs are consistently bicontinuous, triply periodic minimal surface-like microstructures; however, a small but notable gap to the translation bound remains.
\end{abstract}




\end{frontmatter}



\section{Introduction}

Modern energy, structural, and biomedical technologies increasingly demand materials that \emph{simultaneously} transport heat, charge, or fluid while remaining mechanically robust \cite{zhao2025, mackanic2019, dias2012}. Two-phase composites provide a natural platform for achieving such multifunctionality by combining complementary constituents and shaping their spatial arrangement \cite{torquato2004}. In anion-exchange membranes, for example, ionic conductivity must be high while mechanical integrity is maintained under hydration, swelling, and operating loads~\cite{macbeth2025}. Related multifunctional trade-offs arise in lightweight components for transportation and aerospace applications, and in biomaterials where bicontinuous architectures can jointly enhance permeability and stiffness for bone regeneration~\cite{hayashi2023}. Such microstructures can be fabricated directly as mechanical metamaterials~\cite{panetta2015,doyle2025,li2025}, emerge via self-assembly in block copolymers~\cite{fredrickson2006}, or occur in natural photonic composites such as butterfly wing scales~\cite{gan2016}. These examples motivate a central question: \emph{what combinations of effective transport and mechanical properties are fundamentally achievable, and what microstructures attain (or approach) these limits?}

Over the past two decades, topology optimization has become a powerful tool for exploring the enormous design space of periodic composites and for identifying high-performing multifunctional architectures~\cite{bendsoe1988,rozvany1992,wu2021,vitorino2025}. A recurring outcome of these studies is the prominence of bicontinuous geometries resembling triply periodic minimal surfaces (TPMS), which often exhibit near-optimal combinations of transport and stiffness and, in certain settings, attain known performance bounds~\cite{torquato2004,doyle2025,li2025,jung2005,challis2008}. In particular, TPMS such as the Schwarz primitive (P) surface have been shown to provide optimal trade-offs between conductivity and bulk modulus~\cite{torquato2004,challis2008,hyun2002}. These computational discoveries complement the rigorous theory of bounds for effective properties, which not only delineates the feasible region in property space but also guides the interpretation of numerically optimized microstructures.

Among available bounding techniques, the \textit{translation} method \cite{lurie1984,lurie1986,murat1997} has been especially successful in producing tight, general bounds for coupled effective properties. In three dimensions, the translation method yields the sharpest known rigorous bounds for conductivity and bulk modulus in isotropic two-phase composites~\cite{gibiansky96}, by building on earlier cross-property developments and related variational constructions~\cite{gibiansky95,cherkaev92}. In contrast, despite its importance for applications where one phase is mechanically stiff while the other is highly conducting, a comparably general translation-method bound for the conductivity--shear-modulus trade-off in three dimensions has remained unavailable. The difficulty extends to other three-dimensional cross-property problems, with Milton identifying improved bulk--shear-modulus bounds using the translation method as an open challenge~\cite{milton2021openproblems}. The best-known general bounds in the 3D conductivity-shear modulus  setting are due to Berryman and Milton~\cite{berryman88,milton1982}, derived via expansions in microstructural parameters. Such bounds are often looser than translation-method bounds available in two dimensions~\cite{gibiansky95,cherkaev92}, and also in the 3D conductivity--bulk-modulus case~\cite{gibiansky96}. This suggests that tighter 3D conductivity--shear-modulus bounds may be attainable using the translation method.

In this work, we address the longstanding problem of obtaining translation-method cross-property bounds for conductivity and shear modulus in three-dimensional isotropic two-phase composites. Building on the analytical framework developed for related cross-property problems~\cite{gibiansky96,gibiansky95}, we reformulate the quasiconvexity conditions as semidefinite constraints and exploit rotational symmetry to reduce the form of the translation matrix. The resulting bounds are sharper than the previously available Berryman--Milton bounds~\cite{berryman88}. We then perform large-scale topology optimization to search for microstructures that approach the bounds. To model macroscopic isotropy from the periodic unit cells employed in our search, we interpret each homogenized cell as an anisotropic grain and use the self-consistent effective-medium approximation (EMA)~\cite{hill1965,stroud1975,markov2000,han2021} to estimate the response of a statistically isotropic aggregate of randomly oriented grains. Similar self-consistent schemes have shown close agreement with explicit calculations of random polycrystalline aggregates~\cite{counts2008}, supporting comparison of arbitrary periodic structures with the isotropic bounds. Together, the proposed numerical translation framework and the accompanying optimization study provide (i) the tightest conductivity--shear-modulus bounds currently available for isotropic two-phase composites and (ii) a systematic map of realizable microstructures relative to these limits.

Importantly, the underlying construction is not
intrinsically restricted to three spatial dimensions
or to conductivity--shear-modulus bounds. The same
method can, in principle, be extended to higher
dimensions and to other combinations of effective
properties. The present conductivity--shear
problem therefore serves as a first nontrivial
demonstration of a potentially much broader framework for deriving cross-property bounds in heterogeneous materials using the translation method.

\section{Translation Method for $\sigma$--$\mu$ Bounds}
\label{sec:translation_sigma_mu}

We consider a macroscopically isotropic composite composed of two
isotropic phases. Phase $i\in\{1,2\}$ occupies volume fraction $f_i$,
with $f_1+f_2=1$, and is characterized by conductivity $\sigma_i$,
bulk modulus $\kappa_i$, and shear modulus $\mu_i$. The spatial
arrangement of the phases is described by an indicator function
$\rho(r)\in\{0,1\}$, which equals one in phase 1 and zero in phase 2.
We denote the homogenized conductivity, bulk modulus, and shear
modulus by $\sigma^\star$, $\kappa^\star$, and $\mu^\star$,
respectively. Our objective is to derive bounds on the combinations
of $\sigma^\star$ and $\mu^\star$ attainable by any microstructure
with the prescribed phase properties and volume fractions.

To obtain cross-property bounds on conductivity and shear
modulus, we follow the translation method by combining
the electrical and mechanical energies of the composite
into a single quadratic form. This allows the introduction
of a translation matrix that couples the two responses,
providing bounds on their joint effective properties
rather than on each property independently. Both
conductivity and elasticity admit two variational
formulations: one expressed in terms of electric fields
or strains, and the other in terms of currents or
stresses, respectively. Combining these choices gives
four distinct formulations,
$\diamond\in\{\epsilon E,\epsilon J,\tau E,\tau J\}$,
which can provide complementary bounds on different
portions of the admissible $(\sigma^\star,\mu^\star)$
region.

Leaving the details to \ref{app:energies}, for conductivity, the macroscopic electric field and
current are three-dimensional vectors related by the
effective conductivity tensor. To characterize the
full tensorial response, we therefore consider three
mutually orthogonal macroscopic loadings, spanning the
three-dimensional space of possible field directions.
We denote the corresponding local electric and current
fields by $e_n(r)$ and $j_n(r)$, respectively, for
$n=1,2,3$, and assemble them into the $3\times3$
matrices
$\underline{E}(r)=(e_1(r)\ e_2(r)\ e_3(r))$ and
$\underline{J}(r)=(j_1(r)\ j_2(r)\ j_3(r))$.

For elasticity, the strain and stress are symmetric
$3\times3$ tensors, each with six independent components.
These separate into one hydrostatic component, associated
with changes in volume, and five deviatoric components,
associated with shear deformations. Since our
objective is to bound the shear modulus, we consider
five mutually orthogonal macroscopic deviatoric loadings
that span the symmetric traceless subspace. We denote
the corresponding local strain and stress tensors by
$\underline{\epsilon}_n(r)$ and
$\underline{\tau}_n(r)$, respectively, for
$n=1,\ldots,5$.

As detailed in \ref{app:basis} and
\ref{app:energies}, we represent these fields using
a Frobenius-orthonormal basis for $3\times3$ matrices,
which separates the hydrostatic, symmetric deviatoric,
and antisymmetric subspaces. In this basis, the electrical
response matrices $\underline{E}(r)$ and
$\underline{J}(r)$ become nine-dimensional vectors
$\boldsymbol{E}(r)$ and $\boldsymbol{J}(r)$, while each
symmetric strain or stress tensor becomes a
six-dimensional vector $\boldsymbol{\epsilon}_n(r)$
or $\boldsymbol{\tau}_n(r)$. The local isotropic
stiffness and conductivity matrices then take the
diagonal forms
\begin{equation}
\boldsymbol{C}(r)
=\mathrm{diag}(3\kappa(r),2\mu(r)\boldsymbol{I}_5),
\qquad
\boldsymbol{\Sigma}(r)=\sigma(r)\boldsymbol{I}_9,
\label{eq:C_block_iso}
\end{equation}
where $\boldsymbol{I}_n$ is the $n\times n$ identity.
The spatially dependent properties
$x\in\{\kappa,\mu,\sigma\}$ take their corresponding
phase values according to
$x(r)=x_1\rho(r)+x_2[1-\rho(r)]$.
The stress and current formulations involve the
inverse property matrices $\boldsymbol{C}(r)^{-1}$
and $\boldsymbol{\Sigma}(r)^{-1}$, respectively.

The five mechanical loadings contribute $5\times6=30$
field components, while the three electrical loadings
contribute nine, giving a combined 39-dimensional
field space. For each of the four formulations
$\diamond\in\{\epsilon E,\epsilon J,\tau E,\tau J\}$,
we define the stacked field vectors
\begin{align}
\boldsymbol{\nu}_{\epsilon E}(r)
&=
\begin{pmatrix}
\boldsymbol{\epsilon}_1(r)^T & \cdots &
\boldsymbol{\epsilon}_5(r)^T & \boldsymbol{E}(r)^T
\end{pmatrix}^T,
&
\boldsymbol{\nu}_{\epsilon J}(r)
&=
\begin{pmatrix}
\boldsymbol{\epsilon}_1(r)^T & \cdots &
\boldsymbol{\epsilon}_5(r)^T & \boldsymbol{J}(r)^T
\end{pmatrix}^T,
\notag\\
\boldsymbol{\nu}_{\tau E}(r)
&=
\begin{pmatrix}
\boldsymbol{\tau}_1(r)^T & \cdots &
\boldsymbol{\tau}_5(r)^T & \boldsymbol{E}(r)^T
\end{pmatrix}^T,
&
\boldsymbol{\nu}_{\tau J}(r)
&=
\begin{pmatrix}
\boldsymbol{\tau}_1(r)^T & \cdots &
\boldsymbol{\tau}_5(r)^T & \boldsymbol{J}(r)^T
\end{pmatrix}^T,
\label{eq:v_defs_clean}
\end{align}
with corresponding $39\times39$ local property matrices
\begin{align}
\boldsymbol{D}_{\epsilon E}(r)
&=\left(\boldsymbol{I}_5\otimes\boldsymbol{C}(r)\right)
\oplus\boldsymbol{\Sigma}(r),
&
\boldsymbol{D}_{\epsilon J}(r)
&=\left(\boldsymbol{I}_5\otimes\boldsymbol{C}(r)\right)
\oplus\boldsymbol{\Sigma}(r)^{-1},
\notag\\
\boldsymbol{D}_{\tau E}(r)
&=\left(\boldsymbol{I}_5\otimes\boldsymbol{C}(r)^{-1}\right)
\oplus\boldsymbol{\Sigma}(r),
&
\boldsymbol{D}_{\tau J}(r)
&=\left(\boldsymbol{I}_5\otimes\boldsymbol{C}(r)^{-1}\right)
\oplus\boldsymbol{\Sigma}(r)^{-1},
\label{eq:D_defs_clean}
\end{align}
where $\oplus$ denotes a direct sum and $\otimes$ denotes a direct product. Each coupled local energy can then be written as the quadratic form
\begin{equation}
W_\diamond(r)
=\boldsymbol{\nu}_\diamond(r)^T
\boldsymbol{D}_\diamond(r)
\boldsymbol{\nu}_\diamond(r).
\label{eq:mixed energies reduced_clean}
\end{equation}
We denote the constant property matrix of phase $i$ by
$\boldsymbol{D}_\diamond^i$, obtained by substituting the corresponding phase properties into Eq.~\eqref{eq:D_defs_clean}. The macroscopic effective matrix $\boldsymbol{D}_\diamond^\star$ is defined analogously using $\sigma^\star$, $\kappa^\star$, and $\mu^\star$.

The translation method introduces a constant, symmetric matrix
$\boldsymbol{T}$ that modifies the local quadratic energies in a way which bounds the effective properties.
By optimizing the entries of $\boldsymbol{T}$ subject to
appropriate constraints, increasingly restrictive bounds
on the effective properties can be obtained. These bounds take a particularly compact form using the
scalar Y-transform~\cite{milton1991field}.
For a scalar effective property $x^\star$, phase properties
$x_1$ and $x_2$, and volume fractions $f_1$ and $f_2$,
the transform is defined as
\begin{equation}
y(x^\star;x_1,x_2,f_1)
=-f_1x_2-f_2x_1
-\frac{f_1f_2(x_1-x_2)^2}
{x^\star-f_1x_1-f_2x_2}.
\label{eq:y transform}
\end{equation}
When the phase properties and volume fractions are fixed
by context, we use the abbreviated notation $y(x^\star)$. Additional properties
of the transform are given in \ref{app:Y-transform}. Following~\cite{gibiansky96,gibiansky95},
any admissible translation matrix $\boldsymbol{T}$ yields
the bound
\begin{equation}
y(\boldsymbol{D}_\diamond^\star)
+\boldsymbol{T}\succeq0,
\label{eq:translation_bound_main_clean}
\end{equation}
provided that the phase-wise constraints
\begin{equation}
\boldsymbol{D}_\diamond^i-\boldsymbol{T}\succeq0,
\qquad i=1,2,
\label{eq:phase_constraints_clean}
\end{equation}
and the quasiconvexity condition
\begin{equation}
\left\langle
\boldsymbol{\nu}_\diamond^T
\boldsymbol{T}\boldsymbol{\nu}_\diamond
\right\rangle
\geq
\left\langle\boldsymbol{\nu}_\diamond\right\rangle^T
\boldsymbol{T}
\left\langle\boldsymbol{\nu}_\diamond\right\rangle
\label{eq:quasi_convexity_original}
\end{equation}
are satisfied for all compatible fields $\boldsymbol{\nu}_{\diamond}(r)$. Here,
$\boldsymbol{A}\succeq0$ denotes a positive-semidefinite
matrix,
$\langle\cdot\rangle=V^{-1}\int_V dr\,(\cdot)$
denotes the spatial average, and
$y(\boldsymbol{D}_\diamond^\star)$ is understood
entry-wise, with the appropriate phase properties used
for each entry. The main difficulty in optimizing the translation bound
Eq.~\eqref{eq:translation_bound_main_clean} is enforcing
the quasiconvexity condition
Eq.~\eqref{eq:quasi_convexity_original} for arbitrary
admissible microscopic fields. We show that this condition
can be reformulated as a family of positive-semidefinite
constraints, allowing the translation method to be treated
using semidefinite optimization techniques.

Fourier transforming Eq.~\eqref{eq:quasi_convexity_original}
gives
\begin{equation}
    \sum_{k\neq0}
    \tilde{\boldsymbol{\nu}}_\diamond(k)^\dagger
    \boldsymbol{T}
    \tilde{\boldsymbol{\nu}}_\diamond(k)
    \geq0,
    \label{eq:quasi-convexity fourier}
\end{equation}
where $\tilde{\boldsymbol{\nu}}_\diamond(k)$ denotes
the Fourier amplitude of the coupled field at wave vector
$k$, and $\dagger$ denotes the conjugate transpose.
Since the amplitudes of distinct nonzero Fourier modes
can be varied independently, the condition must hold
for each mode separately:
\begin{equation}
    \tilde{\boldsymbol{\nu}}_\diamond(k)^\dagger
    \boldsymbol{T}
    \tilde{\boldsymbol{\nu}}_\diamond(k)
    \geq0,
    \qquad \text{for all }\,k\neq0.
    \label{eq:modewise_qc}
\end{equation}
Although requiring $\boldsymbol{T}\succeq0$
would guarantee this inequality, such a restriction
would produce only trivial bounds. The important
observation is that the microscopic fields are
not arbitrary: they must satisfy their respective
compatibility conditions Eq.~\eqref{eq:compatibility_electric}, \eqref{eq:compatibility_elastic}.
These differential constraints restrict the possible
Fourier amplitudes to a lower-dimensional subspace.

To characterize this subspace, consider any
nonzero wave vector $k$ and choose a rotation
$\underline{R}\in\mathrm{SO}(3)$ such that
\begin{equation}
    k=|k|\,\underline{R}\hat{x},
    \qquad
    \hat{x}=(1,0,0)^T.
    \label{eq:k_rotation_and_x_hat}
\end{equation}
As detailed in \ref{app:rot_dep_decomp_of_field},
the differential field constraints allow us to
express every admissible Fourier amplitude as
\begin{equation}
    \tilde{\boldsymbol{\nu}}_\diamond(k)
    =
    \boldsymbol{U}(\underline{R})
    \boldsymbol{S}_\diamond
    \tilde{\boldsymbol{\nu}}_{0,\diamond}(k),
    \label{eq:rotation_indexed_fourier_field}
\end{equation}
where $\boldsymbol{U}(\underline{R})\in\mathbb{R}^{39\times39}$ is a representation of $\mathrm{SO}(3)$, and $\boldsymbol{S}_\diamond$
is a fixed constant matrix mapping the unconstrained
reference amplitudes $\tilde{\boldsymbol{\nu}}_{0,\diamond}$
associated with the direction $\hat{x}$ into the full field
space. The reference amplitude vector
$\tilde{\boldsymbol{\nu}}_{0,\diamond}$ has dimension 18
for $\diamond=\epsilon E,\tau E$ and 21 for
$\diamond=\epsilon J,\tau J$. Substituting this representation into
Eq.~\eqref{eq:modewise_qc}, and requiring the
inequality to hold for every reference amplitude,
gives the condition
\begin{equation}
    \boldsymbol{S}_\diamond^T
    \boldsymbol{U}(\underline{R})^T
    \boldsymbol{T}
    \boldsymbol{U}(\underline{R})
    \boldsymbol{S}_\diamond
    \succeq0,
    \qquad
    \text{for all }\underline{R}\in\mathrm{SO}(3).
    \label{eq:quasi-convexity psd form}
\end{equation}
Thus, quasiconvexity has been reformulated as a
family of standard positive-semidefinite constraints. Although this representation is now suitable for
numerical optimization, it still involves infinitely
many constraints indexed by rotations. In
\ref{app:old_numerical_approach},
we describe a numerical approach to solving the bounds based on sampling a finite set of rotations.

We can substantially simplify the translation bound by exploiting rotational symmetry. As shown in \ref{app:bound_invce_under_rot_of_T}, if a translation matrix $\boldsymbol{T}$ satisfies the phase and quasiconvexity constraints, then its rotated counterpart $\boldsymbol{U}(\underline{R})^T\boldsymbol{T}\boldsymbol{U}(\underline{R})$ is also admissible and yields the same macroscopic bound. This motivates restricting our search to rotationally invariant translation matrices satisfying
\begin{equation}
    \boldsymbol{U}(\underline{R})^T
    \boldsymbol{T}
    \boldsymbol{U}(\underline{R})
    =\boldsymbol{T},
    \qquad \text{for all }\underline{R}\in\mathrm{SO}(3).
    \label{eq:invt_T_condition}
\end{equation}
Although this restriction is not required for admissibility, it leads to a considerable simplification of the quasiconvexity condition. In particular, Eq.~\eqref{eq:quasi-convexity psd form} reduces exactly to
\begin{equation}
    \boldsymbol{S}_\diamond^T
    \boldsymbol{T}
    \boldsymbol{S}_\diamond
    \succeq0.
    \label{eq:quasi-convexity_rot_invt}
\end{equation}
Thus, the infinite family of rotation-indexed constraints is replaced by a single positive-semidefinite constraint. We assess the effect of this symmetry restriction on the tightness of the bounds numerically in \ref{app:sym_red_bounds}, finding agreement with the unrestricted formulation to within numerical tolerance.

Rotational invariance also greatly reduces the number of independent parameters in $\boldsymbol{T}$. To determine its general form, we exploit the representation theory of $\mathrm{SO}(3)$. As detailed in \ref{app:representation_theory}, there exists a constant orthogonal matrix $\boldsymbol{Q}_{\mathrm{CG}}$ that decomposes $\boldsymbol{U}(\underline{R})$ into irreducible rotational sectors:
\begin{equation}
    \boldsymbol{U}_{\mathrm{CG}}(\underline{R})
    =  (\boldsymbol{I}_2\otimes\boldsymbol{F}_0(\underline{R}))
    \oplus
    (\boldsymbol{I}_2\otimes\boldsymbol{F}_1(\underline{R}))\oplus(\boldsymbol{I}_3\otimes\boldsymbol{F}_2(\underline{R}))
    \oplus\boldsymbol{F}_3(\underline{R})
    \oplus\boldsymbol{F}_4(\underline{R}),
    \label{eq:U_CG}
\end{equation}
where $\boldsymbol{F}_\ell(\underline{R})$ denotes the $(2\ell+1)$-dimensional irreducible representation of $\mathrm{SO}(3)$, and we define for any matrix $\boldsymbol{A}$ its transformed counterpart as  $\boldsymbol{A}_{\mathrm{CG}}=\boldsymbol{Q}_{\mathrm{CG}}^T\boldsymbol{A}\boldsymbol{Q}_{\mathrm{CG}}$. Applying Schur's lemma, the most general symmetric translation matrix satisfying Eq.~\eqref{eq:invt_T_condition} takes the form
\begin{equation}
    \boldsymbol{T}_{\mathrm{CG}}
    =
    \boldsymbol{t}_0
    \oplus(\boldsymbol{t}_1\otimes\boldsymbol{I}_3)
    \oplus(\boldsymbol{t}_2\otimes\boldsymbol{I}_5)
    \oplus t_3\boldsymbol{I}_7
    \oplus t_4\boldsymbol{I}_9,
    \label{eq:trans_mtx_gen_form}
\end{equation}
where $\boldsymbol{t}_0$ and $\boldsymbol{t}_1$ are symmetric $2\times2$ matrices, $\boldsymbol{t}_2$ is a symmetric $3\times3$ matrix, and $t_3,t_4$ are scalars. We have therefore reduced the number of independent translation parameters from 780 to just 14, which we collect into a vector $\boldsymbol{t}\in\mathbb{R}^{14}$. The transformation $\boldsymbol{Q}_{\mathrm{CG}}$ also preserves the hydrostatic elastic, deviatoric elastic, and electrical sectors of the property matrices, whose transformed forms are given in \ref{app:representation_theory}. 

The admissible set of translation parameters becomes
\begin{equation}
\mathcal{S}_{\boldsymbol{t},\diamond}
=\big\{\boldsymbol{t}:\boldsymbol{D}_{\mathrm{CG},\diamond}^i
-\boldsymbol{T}_{\mathrm{CG}}(\boldsymbol{t})
\succeq0,\text{ for } i=1,2,\ \text{and }\boldsymbol{S}_\diamond^T
\boldsymbol{Q}_{\mathrm{CG}}
\boldsymbol{T}_{\mathrm{CG}}(\boldsymbol{t})
\boldsymbol{Q}_{\mathrm{CG}}^T
\boldsymbol{S}_\diamond\succeq0
\big\}.
\label{eq:reduced_feasible_set}
\end{equation}
The corresponding macroscopic bound is
\begin{equation}
    y(\boldsymbol{D}_{\mathrm{CG},\diamond}^\star)
    +\boldsymbol{T}_{\mathrm{CG}}(\boldsymbol{t})
    \succeq0,
    \qquad
    \boldsymbol{t}\in\mathcal{S}_{\boldsymbol{t},\diamond}.
    \label{eq:CG_bounds}
\end{equation}
The bound and phase constraints are block diagonal and involve separate irreducible sectors of the translation matrix. By contrast, the quasiconvexity condition, although greatly simplified by rotational symmetry, still involves nontrivial linear combinations of the translation parameters. Rather than attempting to solve these constraints analytically, we can use an efficient numerical procedure described in \ref{app:sym_red_bounds} to determine the bound.

We provide one further simplification. As described in \ref{app:sym_red_bounds}, numerical comparisons of the full symmetry-reduced bounds over a wide range of material parameters reveal that the tightest $\sigma^\star$--$\mu^\star$ bound is consistently determined by the $\ell=0$ sector of Eq.~\eqref{eq:CG_bounds}. This sector gives the $2\times2$ inequality
\begin{equation}
    \begin{pmatrix}
        y(d^\star_{\mathrm{dev},\diamond}) & 0\\
        0 & y(d^\star_{\mathrm{elec},\diamond})
    \end{pmatrix}
    +\boldsymbol{t}_0\succeq0,
    \label{eq:reduced_CG_bound}
\end{equation}
subject to the full phase and quasiconvexity constraints $\mathcal{S}_{\boldsymbol{t},\diamond}$. Here, the effective property coefficients depend on the formulation $\diamond$ according to
\begin{equation}
\begin{aligned}
d^\star_{\mathrm{dev},\diamond}
&=
\begin{cases}
2\mu^\star, & \diamond=\epsilon E,\epsilon J,\\
1/(2\mu^\star), & \diamond=\tau E,\tau J,
\end{cases}\\
d^\star_{\mathrm{elec},\diamond}
&=
\begin{cases}
\sigma^\star, & \diamond=\epsilon E,\tau E,\\
1/\sigma^\star, & \diamond=\epsilon J,\tau J.
\end{cases}
\end{aligned}
\end{equation}

To determine the bound, we consider the two-dimensional unit vector
$\boldsymbol{u}=(
    \sqrt{\theta} \ \sqrt{1-\theta}
)^T$,
where $\theta\in[0,1]$, and define
\begin{equation}
\gamma_\diamond(\theta)
=-\min_{\boldsymbol{t}\in
\mathcal{S}_{\boldsymbol{t},\diamond}}
\left(
\theta t_{0,1}
+(1-\theta)t_{0,2}
+2\sqrt{\theta(1-\theta)}t_{0,3}
\right).
\label{eq:gamma_main_txt}
\end{equation}
Here, $t_{0,1}$, $t_{0,2}$, and $t_{0,3}$ are the three independent entries of the symmetric matrix $\boldsymbol{t}_0$. For fixed $\theta$, the objective is linear in the translation parameters, while the phase and quasiconvexity constraints are positive-semidefinite matrix inequalities. Thus, the minimization is a convex semidefinite program, which we solve numerically using CVXPY~\cite{diamond2016cvxpy}.

As derived in \ref{app:derivation_of_parametric_bounds}, the boundary of the resulting translation-method bound admits the parametric representation
\begin{equation}
\boxed{
\begin{aligned}
    y(d^\star_{\mathrm{dev},\diamond})
    &=-t_{0,1,\diamond}^*(\theta)
    -t_{0,3,\diamond}^*(\theta)
    \sqrt{\frac{1-\theta}{\theta}},\\[4pt]
    y(d^\star_{\mathrm{elec},\diamond})
    &=-t_{0,2,\diamond}^*(\theta)
    -t_{0,3,\diamond}^*(\theta)
    \sqrt{\frac{\theta}{1-\theta}},
\end{aligned}
\quad\theta\in(0,1),
}
\label{eq:parametric_bound}
\end{equation}
where $t_{0,i,\diamond}^*(\theta)$ are the optimal translation parameters obtained from Eq.~\eqref{eq:gamma_main_txt}. At the endpoints, the limiting boundary lines are given by
\begin{equation}
    \begin{aligned}
    y(d^\star_{\mathrm{elec},\diamond})
    &=\gamma_\diamond(0),\\
    y(d^\star_{\mathrm{dev},\diamond})
    &=\gamma_\diamond(1).
    \end{aligned}
    \label{eq:parametric_bound_endpoints}
\end{equation}
Equations~\eqref{eq:parametric_bound} and \eqref{eq:parametric_bound_endpoints} describe the boundary in the transformed property plane. The actual bound consists of the region on or above and to the right of this boundary.

The corresponding values of $\sigma^\star$ and $\mu^\star$ are recovered using the inverse Y-transform. We repeat this construction for all four formulations $\diamond\in\{\epsilon E,\epsilon J,\tau E,\tau J\}$ and combine the resulting bounds with the individual Hashin--Shtrikman bounds. At each conductivity $\sigma^\star$, the final lower bound on $\mu^\star$ is the largest of the lower bounds, while the final upper bound is the smallest of the upper bounds. Different formulations may therefore determine different portions of the final boundary.

\section{Results}

\begin{figure}[!tb]
    \centering
    \includegraphics[width=0.93\linewidth]{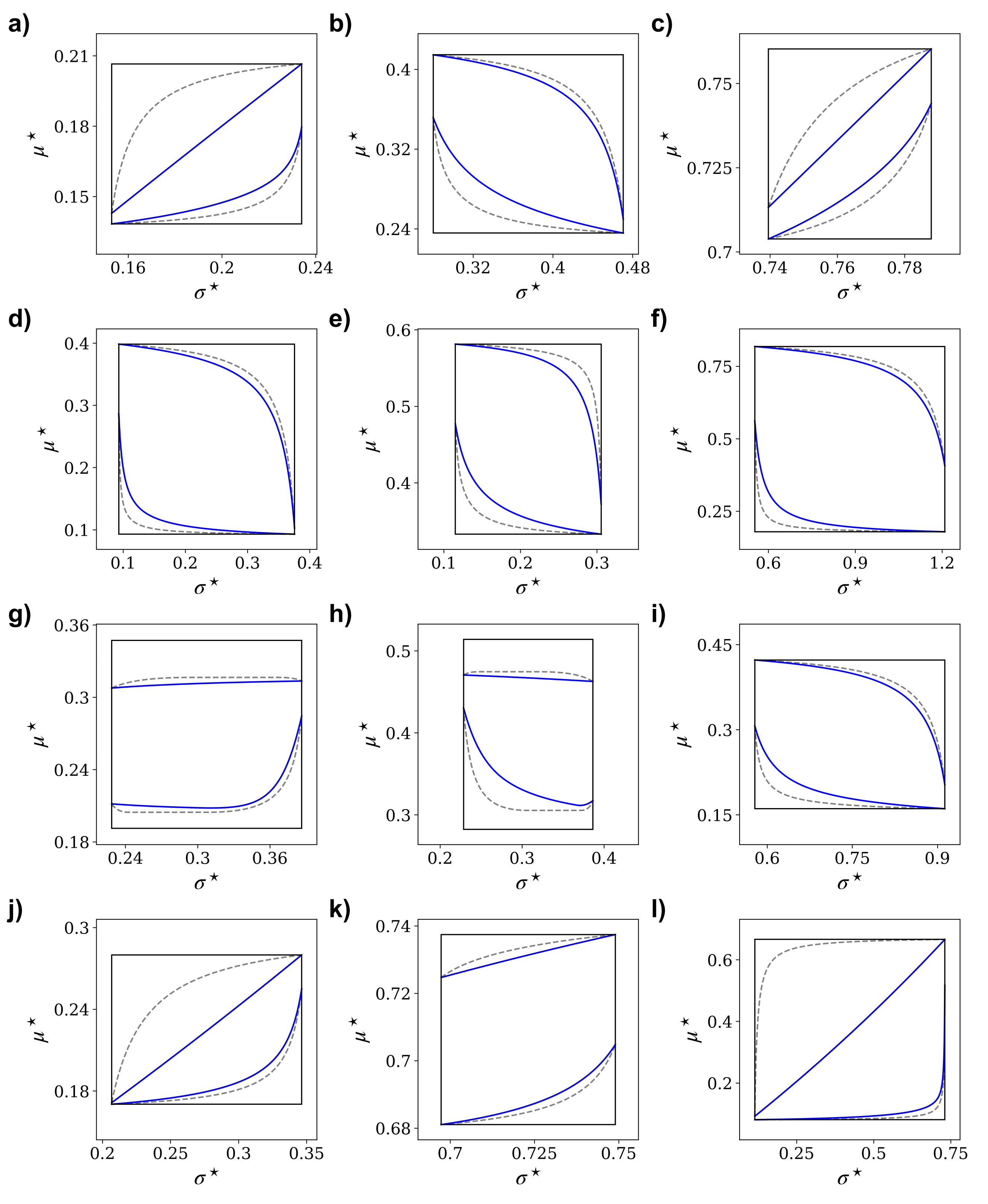}
    \caption{Comparison of cross-property bounds in the $(\sigma^\star,\mu^\star)$ plane for various values of the phase parameters. Solid blue: translation bounds computed with Eq.~\eqref{eq:parametric_bound}, intersected over all four formulations $\diamond\in\{\epsilon E,\epsilon J, \tau E, \tau J\}$. Dashed gray: Berryman--Milton bounds~\cite{berryman88}. Black: individual Hashin--Shtrikman bounds. Writing $\sigma=(\sigma_1,\sigma_2)$, $\kappa=(\kappa_1,\kappa_2)$, and $\mu=(\mu_1,\mu_2)$, with $f=f_1$, the parameters are:
(a) $\sigma=(1,0.1)$, $\kappa=(1,0.1)$, $\mu=(1,0.1)$, $f=0.2$;
(b) $\sigma=(0.1,1)$, $\kappa=(1,0.1)$, $\mu=(1,0.1)$, $f=0.5$;
(c) $\sigma=(1,0.3)$, $\kappa=(1,0.3)$, $\mu=(1,0.3)$, $f=0.75$;
(d) $\sigma=(0.03,1)$, $\kappa=(1,0.03)$, $\mu=(1,0.03)$, $f=0.55$;
(e) $\sigma=(0.05,1)$, $\kappa=(10,0.2)$, $\mu=(1,0.1)$, $f=0.65$;
(f) $\sigma=(0.2,3)$, $\kappa=(4,0.7)$, $\mu=(2,0.05)$, $f=0.55$;
(g) $\sigma=(1,0.1)$, $\kappa=(0.05,2)$, $\mu=(1,0.1)$, $f=0.4$;
(h) $\sigma=(0.1,1)$, $\kappa=(0.05,2)$, $\mu=(1,0.1)$, $f=0.6$;
(i) $\sigma=(1.5,0.15)$, $\kappa=(0.3,3)$, $\mu=(0.08,1.6)$, $f=0.65$;
(j) $\sigma=(1,0.1)$, $\kappa=(0.03,0.003)$, $\mu=(1,0.1)$, $f=0.35$;
(k) $\sigma=(1,0.3)$, $\kappa=(30,9)$, $\mu=(1,0.3)$, $f=0.7$;
(l) $\sigma=(1,0.01)$, $\kappa=(1,0.01)$, $\mu=(1,0.01)$, $f=0.8$.}
    \label{fig:translation_bound}
\end{figure}

Figure~\ref{fig:translation_bound} compares our
$\sigma^\star$--$\mu^\star$ translation-method bounds,
Eqs.~\eqref{eq:parametric_bound} and
\eqref{eq:parametric_bound_endpoints}, with the
Berryman--Milton (BM) bounds~\cite{berryman88} for a wide range of phase-property
contrasts, volume fractions, and relative orderings of electrical and
mechanical properties, including cases where the more conductive phase
is either mechanically softer or stiffer. Across all examples,
our translation bounds improve upon the BM bounds,
reducing the area of the admissible $(\sigma^\star,\mu^\star)$ region
by approximately $7$--$51\%$, with an average reduction of $23\%$. We reproduce the BM  bounds in
\ref{app:BM_bounds} for completeness.

\begin{figure}[!tb]
    \centering
    \includegraphics[width=0.99\linewidth]{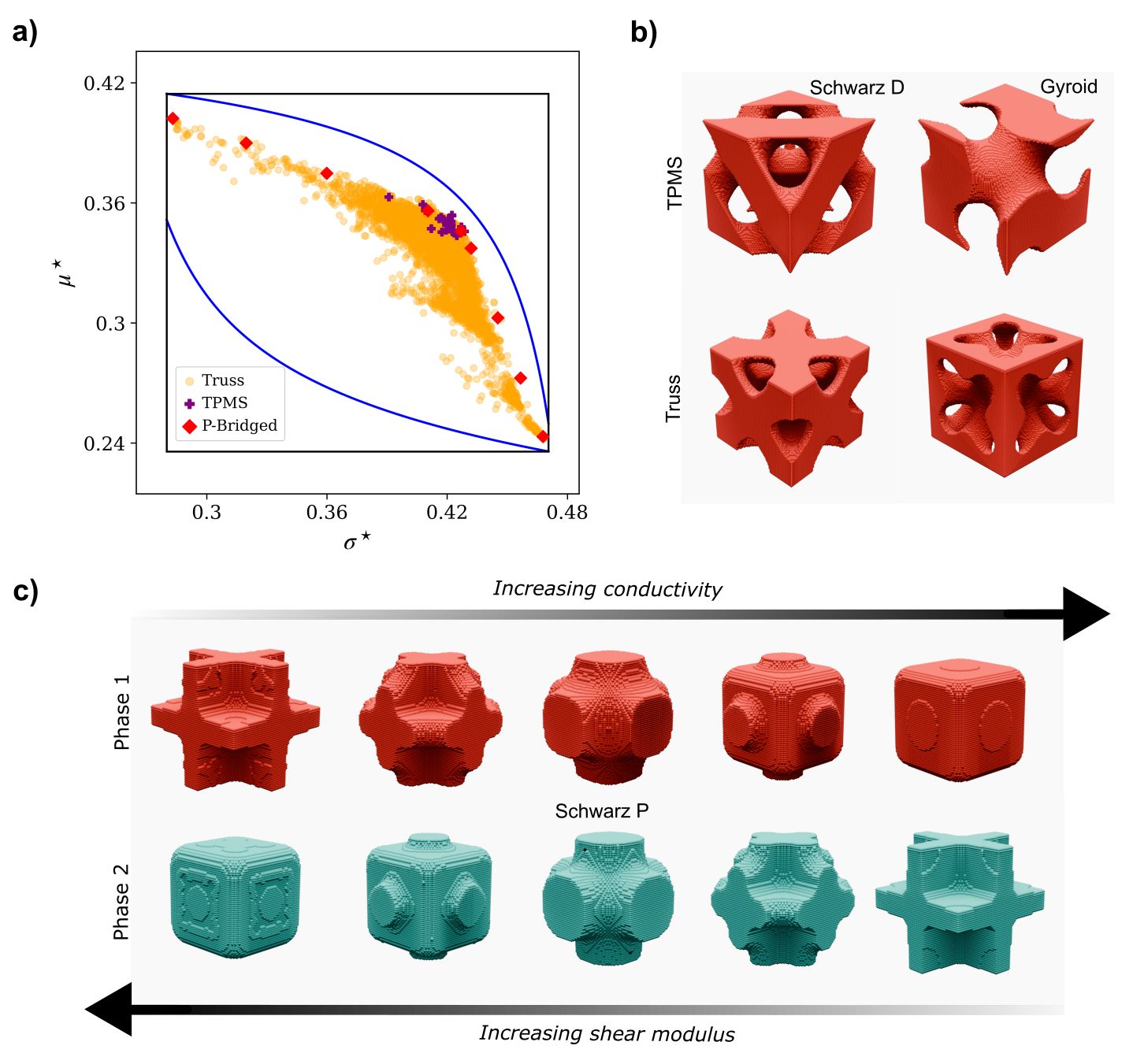}
    \caption{Near-optimal microstructures and bounds. (a) Topology-optimized microstructures compared to the translation bound for phase parameters $\kappa_1=\mu_1=\sigma_2=1.0$ and $\kappa_2=\mu_2=\sigma_1=0.1$ at phase-1 fraction $f_1=0.5$. Blue: translation bound in the $(\sigma^\star,\mu^\star)$ plane. Black: Hashin--Shtrikman bounds for reference. Orange circles: optimized truss initializations. Purple plus symbols: optimized TPMS initializations. Red diamonds: the ``P-bridged'' parametric family, illustrating an approximate empirical upper envelope of the best structures found. (b) Representative near-optimal topology-optimized microstructure unit cells at approximately equal trade-off between conductivity and shear. The visualization shows the phase-1 network (the mechanically stiffer, less conducting phase) within a periodic cubic unit cell; phase 2 occupies the complementary region. Top row: optimized TPMS-derived designs; bottom row: optimized truss-derived designs. All shown examples achieve comparable effective properties, approximately $\sigma^\star\approx0.425$ and $\mu^\star\approx0.350$. (c) The ``P-bridged'' parametric transition observed among the best-performing structures. Top row: phase 1 (high $\kappa,\mu$ / low $\sigma$); bottom row: phase 2 (high $\sigma$ / low $\kappa,\mu$). Progressing left to right, phase 1 morphs from a three-orthogonal-slab network to an approximate Schwarz-P-like bicontinuous morphology and then toward a disconnected cube; simultaneously phase 2 undergoes the complementary transition. For visualization only, the phase-2 field is shifted by $(0.5,0.5,0.5)$ in lattice units; the true phases are complementary and tile the periodic cell.}
    \label{fig:bound_dataset}
\end{figure}

Having established that the translation method yields the sharpest known $\sigma^\star$--$\mu^\star$ bounds, we next ask how closely practical microstructures can approach these bounds. We focus on the case in which the mechanically stiff phase is poorly conducting while the conducting phase is mechanically compliant. This is particularly relevant for polymer membrane-like composites, e.g., hydrated ion-conducting domains embedded in a stiffer matrix~\cite{macbeth2025}. Throughout this section we fix
$\kappa_1=\mu_1=\sigma_2=1.0, \  \kappa_2=\mu_2=\sigma_1=0.1,\ f_1=0.5$,
and we generate candidate unit cell microstructures via 3D periodic topology optimization using the Solid Isotropic Material with Penalization (SIMP) method~\cite{bendsoe1988,rozvany1992}. As described in \ref{app:topopt}, our topology optimization method aims to maximize a weighted sum of the macroscopic properties $\sigma^\star$ and $\mu^\star$, which we obtain by using the effective medium approximation (EMA)~\cite{hill1965,stroud1975,markov2000,han2021} on the calculated unit cell conductivity and stiffness tensors.

A key practical issue is that gradient-based topology optimization is sensitive to initialization. Rather than relying on random starts, we initialize from two diverse and promising families of structures and then locally improve each by SIMP:
(i) a large database of truss-like architectures spanning multiple symmetry classes~\cite{abumualla2024}, and
(ii) a curated set of triply periodic minimal surfaces (TPMS)~\cite{fisher2023}. Most of the runs target a balanced trade-off between conductivity and shear.

Figure~\ref{fig:bound_dataset}(a) compares the resulting optimized designs to the translation bound. Despite thousands of distinct local optima (approximately 5000 truss-based runs and $\sim$30 TPMS-based runs), the best designs remain below the bound, with the largest gap appearing in the high-$\mu^\star$ regime. One possible explanation is that approaching the Hashin--Shtrikman shear-modulus bound may require more complex, potentially hierarchical microstructures~\cite{torquato2002} that are difficult to access through unit-cell SIMP optimization. Nevertheless, both the truss and TPMS families produce designs that approach the bound closely in the intermediate regime corresponding to the most favorable conductivity--shear trade-off.

Figure~\ref{fig:bound_dataset}(b) highlights a small subset of high-performing designs near the apparent “knee” of the dataset envelope. Notably, classical TPMS motifs such as the Schwarz diamond (D), gyroid, and Schwarz primitive (P) (shown in the middle of Figure~\ref{fig:bound_dataset}(c)) appear among the strongest TPMS-based candidates, and their topology-optimized variants perform on par with the best truss-derived solutions. This reinforces the utility of TPMS as compact, physically meaningful initializations for exploring near-optimal cross-property trade-offs.

Beyond individual optima, the large truss-based study
reveals a recurring geometric pattern along the
empirical upper envelope of achieved
$(\sigma^\star,\mu^\star)$ pairs. We observe a
continuous morphological transition from a
load-bearing, three-directional slab network to a
Schwarz-P-like bicontinuous structure, and finally
toward a disconnected, cube-like morphology. We
refer to this pathway as the \emph{P-bridged} family
(Figure~\ref{fig:bound_dataset}(c)). Along this transition, the connectivity of the
conducting phase increases while that of the
mechanically stiff phase decreases, producing a
trade-off between conductivity and shear modulus.
Near the Schwarz-P-like intermediate morphology,
both phases form connected networks, yielding a
particularly favorable compromise and bringing
the structure closest to the translation bound
among the P-bridged samples.

\section{Conclusion}

We have computed translation-method cross-property bounds for conductivity and shear modulus in three-dimensional isotropic two-phase composites, obtaining the sharpest bounds known to date. Large-scale topology optimization further identifies high-performing bicontinuous microstructures, including TPMS-like architectures that approach these bounds. 

Beyond this particular application, the central contribution of this work is a framework for determining translation-method bounds that have resisted analytical treatment due to complicated quasiconvexity conditions. By reformulating quasiconvexity as semidefinite constraints and exploiting rotational symmetry to reduce the number of independent translation parameters, our approach provides a systematic and efficient method for computing bounds in high-dimensional coupled-property problems. The underlying framework is not restricted to three dimensions or to conductivity--shear-modulus bounds, and suggests extensions to higher-dimensional composites, other combinations of material properties, and bounds coupling three or more effective properties simultaneously. These results establish a promising computational direction for extending the translation method beyond problems accessible through conventional analytical approaches.

\section{Acknowledgments}
This work was primarily supported by the Center for Alkaline-Based Energy Solutions (CABES), part of the Energy Frontier Research Center (EFRC) program supported by the U.S. Department of Energy, under grant DE-SC-0019445.

\section{Data Availability}
All code required to reproduce the results in this work is available on GitHub at the link https://github.com/jt577/translation-and-topopt.

\section{Author Contributions}
\textbf{Justin Tahmassebpur}: Conceptualization, Methodology, Software, Validation, Formal analysis, Investigation, Writing - Original Draft. 
\textbf{Tom\'as Arias}: Writing - Review \& Editing, Supervision, Funding Acquisition.

\section{Declaration of generative AI and AI-assisted technologies in the manuscript preparation process}
During the preparation of this work, the authors used OpenAI's ChatGPT to improve the clarity and wording of the manuscript. The authors also used OpenAI Codex to assist with code development. All AI-assisted content was reviewed and validated by the authors, who take full responsibility for the content of the published article.

\appendix

\section{Tensor basis}
\label{app:basis}
We use the following Frobenius-orthonormal basis $\{\underline{a}^{(\alpha)}\}_{\alpha=1}^9$ for $3\times3$ tensors:
\begin{equation}
\begin{aligned}
\underline{a}^{(1)}&=\frac{1}{\sqrt{3}}
\begin{pmatrix}1&0&0\\0&1&0\\0&0&1\end{pmatrix},
&
\underline{a}^{(2)}&=\frac{1}{\sqrt{2}}
\begin{pmatrix}1&0&0\\0&-1&0\\0&0&0\end{pmatrix},
&
\underline{a}^{(3)}&=\frac{1}{\sqrt{6}}
\begin{pmatrix}1&0&0\\0&1&0\\0&0&-2\end{pmatrix},
\\[10pt]
\underline{a}^{(4)}&=\frac{1}{\sqrt{2}}
\begin{pmatrix}0&1&0\\1&0&0\\0&0&0\end{pmatrix},
&
\underline{a}^{(5)}&=\frac{1}{\sqrt{2}}
\begin{pmatrix}0&0&1\\0&0&0\\1&0&0\end{pmatrix},
&
\underline{a}^{(6)}&=\frac{1}{\sqrt{2}}
\begin{pmatrix}0&0&0\\0&0&1\\0&1&0\end{pmatrix},
\\[10pt]
\underline{a}^{(7)}&=\frac{1}{\sqrt{2}}
\begin{pmatrix}0&1&0\\-1&0&0\\0&0&0\end{pmatrix},
&
\underline{a}^{(8)}&=\frac{1}{\sqrt{2}}
\begin{pmatrix}0&0&1\\0&0&0\\-1&0&0\end{pmatrix},
&
\underline{a}^{(9)}&=\frac{1}{\sqrt{2}}
\begin{pmatrix}0&0&0\\0&0&1\\0&-1&0\end{pmatrix}\!,
\end{aligned}
\label{eq:3x3 basis}
\end{equation}
where $\underline{a}^{(1)}$ is hydrostatic, $\underline{a}^{(2)}$--$\underline{a}^{(6)}$ are symmetric deviatoric, and $\underline{a}^{(7)}$--$\underline{a}^{(9)}$ are antisymmetric. Defining the double contraction $\underline{a}:\underline{b}=\sum_{ij}a_{ij}b_{ij}$, any $3\times3$ matrix $\underline{X}$ can thus be identified as a vector $\boldsymbol{X}\in\mathbb{R}^9$ with components obtained by that matrix's projections against the above Frobenius matrices $\underline{a}^{(\alpha)}$,
\begin{equation}
X^{(\alpha)}=\underline{X}:\underline{a}^{(\alpha)},
\label{eq:3x3 matrix decomp}
\end{equation}
and any fourth-order tensor $\underline{Y}$ acting on $3\times3$ matrices can likewise be identified with the $9\times 9$ matrix $\boldsymbol{Y}\in\mathbb{R}^{9\times 9}$ with components
\begin{equation}
Y^{(\alpha\beta)}=\underline{a}^{(\alpha)}:\,\underline{Y}:\,\underline{a}^{(\beta)}.
\label{eq:3x3x3x3 tensor decomp}
\end{equation}

\section{Electrostatic, elastic, and coupled energies}
\label{app:energies}

For conductivity or resistivity, respectively, we determine the material response considering either the spatially-dependent microscopic electric field $e$ or the microscopic current field, $j=\underline{\sigma}\,e$ with compatibility constraints
\begin{equation}
    e=-\nabla\phi, \qquad \nabla\cdot j=0.
    \label{eq:compatibility_electric}
\end{equation}
For brevity, we suppress the spatial dependencies $e(r)$, $j(r)$, etc. For isotropic phases, the conductivity tensor is diagonal, $\underline{\sigma}=\sigma\,\underline{I}_3$. The local energy-dissipation density may then be written as
\begin{equation}
W_{e}=e\cdot \underline{\sigma}\cdot e,
\qquad
W_{j}=j\cdot \underline{\sigma}^{-1}\cdot j,
\label{eq:cond_pointwise_energies}
\end{equation}
where $\underline{\sigma}^{-1}$ is the microscopic resistivity tensor.

To characterize the entire response, we consider three mutually orthogonal macroscopic loadings and assemble the resulting microscopic response fields into the $3\times3$ matrices
\[
\underline{E}=(e_1\ e_2\ e_3),\qquad
\underline{J}=(j_1\ j_2\ j_3),
\]
where the local electric $e_i$ and local current $j_i$ fields are column vectors. The total local energy-dissipation density is the sum of the energies associated with each field $i=1,2,3$, which can be written in the compact form
\begin{equation}
    W_{E}=\mathrm{Tr}(\underline{E}^T\underline{\sigma} \ \underline{E}), \quad W_{J}=\mathrm{Tr}(\underline{J}^T\underline{\sigma}^{-1} \ \underline{J})
\end{equation}
 In the basis Eq.~\eqref{eq:3x3 basis}, the matrices $\underline{E}$ and $\underline{J}$ can be collected into nine dimensional vectors $\boldsymbol{E},\boldsymbol{J}$, giving 
\begin{equation}
W_{E}=\boldsymbol{E}\cdot \boldsymbol{\Sigma}\cdot \boldsymbol{E},
\qquad
W_{J}=\boldsymbol{J}\cdot \boldsymbol{\Sigma}^{-1}\cdot \boldsymbol{J},
\label{eq:multidirectional_cond_energies}
\end{equation}
where $\boldsymbol{\Sigma}=\sigma\,\boldsymbol{I}_9$, with $\sigma$ being the isotropic spatially-dependent local conductivity.  

For elasticity, the situation is naturally more complicated and, here, the utility of the basis Eq.~\eqref{eq:3x3 basis} comes to the fore. First, we define the hydrostatic and deviatoric four-tensor projectors
\begin{equation}
    I^{\mathrm{hydro}}_{ijkl}
=\frac13\delta_{ij}\delta_{kl},
\qquad
I^{\mathrm{dev}}_{ijkl}
=\frac12(\delta_{ik}\delta_{jl}+\delta_{il}\delta_{jk})
-\frac13\delta_{ij}\delta_{kl},
\label{eq:id_hydro_dev}
\end{equation}
so that the isotropic stiffness tensor becomes simply
\begin{equation}
    \underline{C}=3\kappa \underline{I}^{\mathrm{hydro}} + 2\mu\underline{I}^{\mathrm{dev}},
    \label{eq:stiffness_pos_dep}
\end{equation}
where $\kappa$ and $\mu$ are the local isotropic bulk and shear modulus fields, and we again suppress spatial dependencies. To obtain elastic energy densities, we use either a symmetric strain field $\underline{\epsilon}$ or stress field $\underline{\tau}=\underline{C}:\underline{\epsilon}$, satisfying compatibility conditions
\begin{equation}
    \underline{\epsilon}=\tfrac12(\underline{\zeta}+\underline{\zeta}^T),\qquad \nabla\cdot \underline{\tau}=0,
    \label{eq:compatibility_elastic}
\end{equation}
where $\underline{\zeta}=\nabla u$ and $u$ is the displacement vector. The local energy density then may be written as
\begin{equation}
W_{\epsilon}=\underline{\epsilon}:\underline{C}:\underline{\epsilon},
\qquad
W_{\tau}=\underline{\tau}:\underline{C}^{-1}:\underline{\tau}.
\label{eq:elas_pointwise_energies}
\end{equation}
Note that, here and throughout, we use $\underline{\tau}$ to denote the elastic stress tensor in order to avoid confusion with the electric conductivity tensor $\underline{\sigma}$.

To characterize the shear response, we now apply five mutually orthogonal deviatoric loadings, corresponding to the five-dimensional symmetric traceless subspace. By contrast, a bulk-modulus bound would require only a single hydrostatic loading, since the hydrostatic subspace is spanned by the identity tensor alone. We denote the five deviatoric local strain or stress loadings by \(\{\boldsymbol{\epsilon}_n\}_{n=1}^5\) and \(\{\boldsymbol{\tau}_n\}_{n=1}^5\), respectively. These fields are represented in the symmetric part of the basis Eq.~\eqref{eq:3x3 basis}, so that each strain or stress vector has six components: one hydrostatic component and five deviatoric components. We can assemble these fields into the column vectors of matrices
\begin{equation}
    \boldsymbol{\mathcal{E}} = 
    \begin{pmatrix}
        \boldsymbol{\epsilon}_1 & \boldsymbol{\epsilon}_2 & \dots & \boldsymbol{\epsilon}_5
    \end{pmatrix}
    , \qquad
    \boldsymbol{\mathcal{T}} = 
    \begin{pmatrix}
        \boldsymbol{\tau}_1 & \boldsymbol{\tau}_2 & \dots & \boldsymbol{\tau}_5
    \end{pmatrix}.
\end{equation}
Analogously to the above, the total elastic energies are obtained by summing the energy contributions from the five independent loadings, which may be written compactly as:
\begin{equation}
W_{\mathcal{E}}=\mathrm{Tr}\left(\boldsymbol{\mathcal{E}}^T\boldsymbol{C}\boldsymbol{\mathcal{E}}\right),
\qquad
W_{\mathcal{T}}=\mathrm{Tr}\left(\boldsymbol{\mathcal{T}}^T\boldsymbol{C}^{-1}\boldsymbol{\mathcal{T}}\right).
\label{eq:multidirectional_elas_energies}
\end{equation}
In this basis, the local isotropic stiffness tensor is diagonal and given by Eq.~\eqref{eq:C_block_iso},
where the first entry corresponds to the hydrostatic mode and the remaining five entries correspond to the deviatoric modes. Note that \(\boldsymbol{C}\) is a \(6\times6\) matrix, since only the symmetric part of the full \(3\times3\) basis contributes to elasticity.

The translation method~\cite{gibiansky95} calls for the introduction of four coupled energies:
\begin{align}
W_{\epsilon E} &= W_{\mathcal{E}} + W_{E}, &
W_{\epsilon J} &= W_{\mathcal{E}} + W_{J}, \notag\\
W_{\tau E} &= W_{\mathcal{T}} + W_{E}, &
W_{\tau J} &= W_{\mathcal{T}} + W_{J}.
\label{eq:mixed energies full_clean}
\end{align}
Each choice corresponds to a different combination of variational formulations for the elastic and conductive problems, and therefore to a different side of the attainable region in the \((\sigma^\star,\mu^\star)\) plane. Stacking the fields into the vectors $\boldsymbol{\nu}_{\diamond}$ defined by Eq.~\eqref{eq:v_defs_clean}, and assembling the property matrices as Eq.~\eqref{eq:D_defs_clean}, we arrive at the quadratic form for the coupled energies Eq.~\eqref{eq:mixed energies reduced_clean}.

\section{Y-transform properties}
\label{app:Y-transform}

Two well-known useful properties of the transform are its compatibility with inversion and uniform rescaling:
\begin{align}
    y\left(\frac{1}{x^\star}\right)&=y\left(\frac{1}{x^\star};\frac{1}{x_1},\frac{1}{x_2},f_1\right)=\frac{1}{y(x^\star)} \notag \\
    y\left(ax^\star\right)&=y\left(ax^\star;ax_1,ax_2,f_1\right)=ay(x^\star) \notag.
\end{align}
These identities are convenient here because the four coupled formulations contain both direct properties and their inverses.

\section{Rotation-dependent decomposition of $\boldsymbol{\tilde{\nu}}_{\diamond}(k)$}
\label{app:rot_dep_decomp_of_field}

Here we derive the rotation-dependent decomposition of the Fourier field $\boldsymbol{\tilde{\nu}}_{\diamond}(k)$ in Eq.~\eqref{eq:rotation_indexed_fourier_field}. In the following, we also suppress the explicit dependence of the Fourier-transformed fields on $k$ whenever it is unambiguous. 

We first recall that Eq.~\eqref{eq:k_rotation_and_x_hat} expresses wave vector $k=|k|\underline{R}\hat{x}$ for rotation $\underline{R}\in\mathrm{SO(3)}$ and reference direction $\hat{x}$. In real space, the physical fields are constrained according to $e=-\nabla\phi$, $\nabla\cdot j=0$, $\underline{\epsilon}=\frac{1}{2}(\nabla u+(\nabla u)^T)$, and $\nabla\cdot\underline{\tau}=0$. Using $\underline{\zeta}=\nabla u$, we obtain, in Fourier space, the constraints
\begin{equation}
    \tilde{\underline{E}}=-ik\tilde{\phi}^T,\qquad
\tilde{\underline{\zeta}}=ik\tilde{u}^T,\qquad
k\cdot\tilde{\underline{J}}=0,\qquad
k\cdot\tilde{\underline{\tau}}
=\tilde{\underline{\tau}}\cdot k=0,
\end{equation}
where $\tilde{\phi}=(\tilde{\phi}_1 \ \tilde{\phi}_2 \ \tilde{\phi}_3)^T$ holds the three independent electrostatic potential fields, and $\tilde{u}=(\tilde{u}_1 \ \tilde{u}_2 \ \tilde{u}_3)^T$ holds the three components of the displacement field. We may write the Fourier transformed fields as rotated versions of reference matrices with $k\parallel\hat{x}$. Let us define the reference ($k\parallel \hat{x}$) matrices
\begin{equation}
\begin{alignedat}{2}
\tilde{\underline{E}}_0
&=\begin{pmatrix}
\tilde{E}^1_0 & \tilde{E}^2_0 & \tilde{E}^3_0\\
0 & 0 & 0\\
0 & 0 & 0
\end{pmatrix},\qquad
&\tilde{\underline{\zeta}}_0
&=\begin{pmatrix}
\tilde{\zeta}^1_0 & \tilde{\zeta}^2_0 & \tilde{\zeta}^3_0\\
0 & 0 & 0\\
0 & 0 & 0
\end{pmatrix},\\[4pt]
\tilde{\underline{J}}_0
&=\begin{pmatrix}
0 & 0 & 0\\
\tilde{J}^1_0 & \tilde{J}^2_0 & \tilde{J}^3_0\\
\tilde{J}^4_0 & \tilde{J}^5_0 & \tilde{J}^6_0
\end{pmatrix},\qquad
&\tilde{\underline{\tau}}_0
&=\begin{pmatrix}
0 & 0 & 0\\
0 & \tilde{\tau}^1_0 & \tilde{\tau}^2_0\\
0 & \tilde{\tau}^2_0 & \tilde{\tau}^3_0
\end{pmatrix}.
\end{alignedat}
\label{eq:o matrices}
\end{equation}
The nonzero entries of these reference matrices are independent, unconstrained amplitudes. Absorbing the magnitude $|k|$ and, where necessary, the imaginary unit $i$ into these amplitudes, the Fourier fields may be written as
\begin{equation}
\tilde{\underline{E}}
=\underline{R}\,\tilde{\underline{E}}_0 \underline{R}^T,\qquad
\tilde{\underline{\zeta}}
=\underline{R}\,\tilde{\underline{\zeta}}_0 \underline{R}^T,\qquad
\tilde{\underline{J}}
=\underline{R}\,\tilde{\underline{J}}_0 \underline{R}^T,\qquad
\tilde{\underline{\tau}}
=\underline{R}\,\tilde{\underline{\tau}}_0 \underline{R}^T.
\label{eq:rotated_reference_fourier_fields}
\end{equation}
The additional rotations on the right of $\tilde{\underline{E}}_0$, $\tilde{\underline{\zeta}}_0$, and $\tilde{\underline{J}}_0$ preserve their respective patterns of zero entries. Since these rotations are invertible and the nonzero entries are independent, unconstrained amplitudes, they merely redefine the reference amplitudes without changing the space of admissible Fourier fields. 
Thus, an arbitrary Fourier mode is completely specified by a rotation $\underline{R}$, which determines the direction of the wave vector, together with a set of unconstrained reference amplitudes. We collect these amplitudes into the vectors
\begin{equation*}
\begin{alignedat}{2}
\tilde{E}_0
&=\begin{pmatrix}\tilde{E}^1_0 & \tilde{E}^2_0 & \tilde{E}^3_0\end{pmatrix}^T,\qquad
&\tilde{\zeta}_0
&=\begin{pmatrix}\tilde{\zeta}^1_0 & \tilde{\zeta}^2_0 & \tilde{\zeta}^3_0\end{pmatrix}^T,\\
\tilde{J}_0
&=\begin{pmatrix}\tilde{J}^1_0 & \ldots & \tilde{J}^6_0\end{pmatrix}^T,\qquad
&\tilde{\tau}_0
&=\begin{pmatrix}\tilde{\tau}^1_0 & \tilde{\tau}^2_0 & \tilde{\tau}^3_0\end{pmatrix}^T.
\end{alignedat}
\end{equation*}
Expressing the rotated fields in the Frobenius-orthonormal basis, and denoting a generic rotation by $\underline{R}$, Eq.~\eqref{eq:rotated_reference_fourier_fields} can then be written compactly as
\begin{equation}
\begin{aligned}
\tilde{\boldsymbol{E}}
&=\boldsymbol{G}(\underline{R})\,\boldsymbol{L}\,\tilde{E}_0,
&\qquad
\tilde{\boldsymbol{\epsilon}}
&=\boldsymbol{G}_{\mathrm{sym}}(\underline{R})\,\boldsymbol{P}\,\tilde{\zeta}_0,
\\[4pt]
\tilde{\boldsymbol{J}}
&=\boldsymbol{G}(\underline{R})\,\boldsymbol{N}\,\tilde{J}_0,
&\qquad
\tilde{\boldsymbol{\tau}}
&=\boldsymbol{G}_{\mathrm{sym}}(\underline{R})\,\boldsymbol{M}\,\tilde{\tau}_0.
\end{aligned}
\label{eq:fourier fields in basis}
\end{equation}
Here, the rotation-dependent orthogonal matrix $\boldsymbol{G}(\underline{R})\in\mathbb{R}^{9\times9}$ describes the action of $\underline{R}$ in the Frobenius-orthonormal basis and are defined by
\begin{equation}
G^{(\alpha\beta)}(\underline{R})
=\underline{a}^{(\alpha)}:\big(\underline{R} \underline{a}^{(\beta)} \underline{R}^T\big),
\qquad \alpha,\beta=1,\ldots,9.
\label{eq:G matrix defn}
\end{equation}
The reduced orthogonal matrix $\boldsymbol{G}_{\mathrm{sym}}(\underline{R})\in\mathbb{R}^{6\times6}$ consists of the first six rows and columns of $\boldsymbol{G}(\underline{R})$, corresponding to the symmetric part of the tensor basis. The remaining matrices $\boldsymbol{L}\in\mathbb{R}^{9\times3}$, $\boldsymbol{M}\in\mathbb{R}^{6\times3}$, $\boldsymbol{N}\in\mathbb{R}^{9\times6}$, and $\boldsymbol{P}\in\mathbb{R}^{6\times3}$ are constant maps from the independent reference amplitudes to their full basis representations and are defined as
\begin{equation}
\begin{alignedat}{2}
\boldsymbol{L}
&=\begin{pmatrix}
\frac{1}{\sqrt{3}} & 0 & 0\\
\frac{1}{\sqrt{2}} & 0 & 0\\
\frac{1}{\sqrt{6}} & 0 & 0\\
0 & \frac{1}{\sqrt{2}} & 0\\
0 & 0 & \frac{1}{\sqrt{2}}\\
0 & 0 & 0\\
0 & \frac{1}{\sqrt{2}} & 0\\
0 & 0 & \frac{1}{\sqrt{2}}\\
0 & 0 & 0
\end{pmatrix},\qquad
&\boldsymbol{M}
&=\begin{pmatrix}
\frac{1}{\sqrt{3}} & 0 & \frac{1}{\sqrt{3}}\\
-\frac{1}{\sqrt{2}} & 0 & 0\\
\frac{1}{\sqrt{6}} & 0 & -\sqrt{\frac{2}{3}}\\
0 & 0 & 0\\
0 & 0 & 0\\
0 & \sqrt{2} & 0
\end{pmatrix},\\[6pt]
\boldsymbol{N}
&=\begin{pmatrix}
0 & \frac{1}{\sqrt{3}} & 0 & 0 & 0 & \frac{1}{\sqrt{3}}\\
0 & -\frac{1}{\sqrt{2}} & 0 & 0 & 0 & 0\\
0 & \frac{1}{\sqrt{6}} & 0 & 0 & 0 & -\sqrt{\frac{2}{3}}\\
\frac{1}{\sqrt{2}} & 0 & 0 & 0 & 0 & 0\\
0 & 0 & 0 & \frac{1}{\sqrt{2}} & 0 & 0\\
0 & 0 & \frac{1}{\sqrt{2}} & 0 & \frac{1}{\sqrt{2}} & 0\\
-\frac{1}{\sqrt{2}} & 0 & 0 & 0 & 0 & 0\\
0 & 0 & 0 & -\frac{1}{\sqrt{2}} & 0 & 0\\
0 & 0 & \frac{1}{\sqrt{2}} & 0 & -\frac{1}{\sqrt{2}} & 0
\end{pmatrix},\qquad
&\boldsymbol{P}
&=\begin{pmatrix}
\frac{1}{\sqrt{3}} & 0 & 0\\
\frac{1}{\sqrt{2}} & 0 & 0\\
\frac{1}{\sqrt{6}} & 0 & 0\\
0 & \frac{1}{\sqrt{2}} & 0\\
0 & 0 & \frac{1}{\sqrt{2}}\\
0 & 0 & 0
\end{pmatrix}.
\end{alignedat}
\label{eq:LMNP matrices}
\end{equation}
In deriving Eq.\eqref{eq:fourier fields in basis}, we have used the orthogonality of $\boldsymbol{G}$ and the decomposition 
\begin{align}
    \boldsymbol{G}(\underline{R})&=
        \boldsymbol{G}_{\mathrm{sym}}(\underline{R})\oplus \boldsymbol{G}_{\mathrm{asym}}(\underline{R}),
\end{align}
where $\boldsymbol{G}_{\mathrm{asym}}\in\mathbb{R}^{3\times 3}$ is the part of $\boldsymbol{G}(\underline{R})$ involving only the antisymmetric bases. 

We now assemble the five strain fields into the matrix
\begin{equation}
    \tilde{\boldsymbol{\mathcal{E}}}
    =
    \boldsymbol{G}_{\mathrm{sym}}(\underline{R})
    \boldsymbol{P}
    \begin{pmatrix}
        \tilde{\zeta}_{0,1} & \cdots & \tilde{\zeta}_{0,5}
    \end{pmatrix},
    \label{eq:big_eps_fourier}
\end{equation}
where each $\tilde{\zeta}_{0,n}\in\mathbb{C}^3$ contains
the three independent Fourier amplitudes associated with
the $n$th strain field. Since these amplitudes are independent and unconstrained, we may perform an arbitrary invertible linear transformation
among the five loading indices without changing the space
of admissible Fourier fields. Although this would not change the Fourier compatibility and equilibrium conditions, it
allows the rotation of the spatial tensor components
and the loading indices to be expressed together and is an exploitable symmetry of our approach.
In particular, consider the $5\times5$ orthogonal matrix
$\boldsymbol{G}_{\mathrm{dev}}(\underline{R})$, defined as the deviatoric part of $\boldsymbol{G}(\underline{R})$:
\begin{equation}
    G_{\mathrm{dev}}^{(\alpha\beta)}(\underline{R})
    =
    \underline{a}^{(\alpha+1)}:
    \left(
    \underline{R}\underline{a}^{(\beta+1)}
    \underline{R}^T
    \right),
    \qquad \alpha,\beta=1,\ldots,5.
\end{equation}
This matrix represents the rotation of the five-dimensional
space of symmetric deviatoric tensors. Defining the matrix of reference amplitudes as
\begin{equation}
    \boldsymbol{A}_0=
    \begin{pmatrix}
        \tilde{\zeta}_{0,1} & \cdots & \tilde{\zeta}_{0,5}
    \end{pmatrix},
\end{equation}
we can introduce new amplitudes
$\boldsymbol{A}_0'=\boldsymbol{A}_0
\boldsymbol{G}_{\mathrm{dev}}(\underline{R})$.
Using the orthogonality of $\boldsymbol{G}_{\mathrm{dev}}$, we obtain
\begin{equation}
\begin{aligned}
    \tilde{\boldsymbol{\mathcal{E}}}
    &=
    \boldsymbol{G}_{\mathrm{sym}}(\underline{R})
    \boldsymbol{P}\boldsymbol{A}_0\\
    &=
    \boldsymbol{G}_{\mathrm{sym}}(\underline{R})
    \boldsymbol{P}\boldsymbol{A}_0'
    \boldsymbol{G}_{\mathrm{dev}}(\underline{R})^T.
\end{aligned}
\end{equation}
Because the transformation between
$\boldsymbol{A}_0$ and $\boldsymbol{A}_0'$ is invertible,
the two expressions describe exactly the same set of
admissible Fourier fields. We may therefore relabel
the new amplitudes as $\boldsymbol{A}_0$ and write
\begin{equation}
    \tilde{\boldsymbol{\mathcal{E}}}
    =
    \boldsymbol{G}_{\mathrm{sym}}(\underline{R})
    \boldsymbol{P}
    \begin{pmatrix}
        \tilde{\zeta}_{0,1} & \cdots & \tilde{\zeta}_{0,5}
    \end{pmatrix}
    \boldsymbol{G}_{\mathrm{dev}}(\underline{R})^T.
    \label{eq:big_eps_fourier_withF_strain}
\end{equation}
An analogous representation holds for the five stress
fields,
\begin{equation}
    \tilde{\boldsymbol{\mathcal{T}}}
    =
    \boldsymbol{G}_{\mathrm{sym}}(\underline{R})
    \boldsymbol{M}
    \begin{pmatrix}
        \tilde{\tau}_{0,1} & \cdots & \tilde{\tau}_{0,5}
    \end{pmatrix}
    \boldsymbol{G}_{\mathrm{dev}}(\underline{R})^T.
    \label{eq:big_eps_fourier_withF_stress}
\end{equation}

The quasiconvexity inequality in Eq.~\eqref{eq:modewise_qc} is expressed in terms of the stacked Fourier field $\tilde{\boldsymbol{\nu}}_\diamond$, where we have reintroduced the subscript $\diamond$ for clarity. We therefore combine the individual relations in Eq.~\eqref{eq:fourier fields in basis}, \eqref{eq:big_eps_fourier_withF_strain}, and \eqref{eq:big_eps_fourier_withF_stress} into a single linear representation for each $\diamond\in\{\epsilon E,\epsilon J,\tau E,\tau J\}$. To do so, we define the vectors of unconstrained reference amplitudes
\begin{equation}
\begin{alignedat}{2}
\tilde{\nu}_{0,\epsilon E}
&=\begin{pmatrix}\tilde{\zeta}_{0,1}^T & \dots & \tilde{\zeta}_{0,5}^T & \tilde{E}_0^T\end{pmatrix}^T,\qquad
&\tilde{\nu}_{0,\epsilon J}
&=\begin{pmatrix}\tilde{\zeta}_{0,1}^T & \dots & \tilde{\zeta}_{0,5}^T & \tilde{J}_0^T\end{pmatrix}^T,\\
\tilde{\nu}_{0,\tau E}
&=\begin{pmatrix}\tilde{\tau}_{0,1}^T & \dots & \tilde{\tau}_{0,5}^T & \tilde{E}_0^T\end{pmatrix}^T,\qquad
&\tilde{\nu}_{0,\tau J}
&=\begin{pmatrix}\tilde{\tau}_{0,1}^T & \dots & \tilde{\tau}_{0,5}^T & \tilde{J}_0^T\end{pmatrix}^T,
\end{alignedat}
\label{eq:v_o vectors}
\end{equation}
the rotation matrix $\boldsymbol{U}(\underline{R})$ as 
\begin{equation}
    \boldsymbol{U}(\underline{R})=
        (\boldsymbol{G}_{\mathrm{dev}}(\underline{R})
        \otimes
        \boldsymbol{G}_{\mathrm{sym}}(\underline{R}))
        \oplus\boldsymbol{G}(\underline{R}),
\end{equation}
where $\otimes$ denotes a direct product, 
and the field-specific constant block diagonal matrices $\boldsymbol{S}_\diamond$ as
\begin{equation}
\begin{alignedat}{2}
\boldsymbol{S}_{\epsilon E}
&=\underbrace{\boldsymbol{P}\oplus\ldots\oplus\boldsymbol{P}}_{5\ \text{blocks}}\oplus\boldsymbol{L},\qquad
&\boldsymbol{S}_{\epsilon J}
&=\underbrace{\boldsymbol{P}\oplus\ldots\oplus\boldsymbol{P}}_{5\ \text{blocks}}\oplus\boldsymbol{N},\\
\boldsymbol{S}_{\tau E}
&=\underbrace{\boldsymbol{M}\oplus\ldots\oplus\boldsymbol{M}}_{5\ \text{blocks}}\oplus\boldsymbol{L},\qquad
&\boldsymbol{S}_{\tau J}
&=\underbrace{\boldsymbol{M}\oplus\ldots\oplus\boldsymbol{M}}_{5\ \text{blocks}}\oplus\boldsymbol{N}.
\end{alignedat}
\label{eq:S matrices_clean}
\end{equation}
Combining these objects thus yields Eq.~\eqref{eq:rotation_indexed_fourier_field}.

\section{Finding tight bounds without symmetry reduction}
\label{app:old_numerical_approach}

Although the bound Eq.~\eqref{eq:translation_bound_main_clean} subject to constraints Eq.~\eqref{eq:phase_constraints_clean},~\eqref{eq:quasi-convexity psd form} can now be found using standard semidefinite optimization techniques, determining the tightest bound still requires a suitable objective function. This is not entirely straightforward because the quantity of interest is not a single scalar property, but the boundary of the attainable region in the coupled $(\sigma^\star,\mu^\star)$ property space. We therefore introduce a numerical procedure that traces this boundary along prescribed directions and simultaneously optimizes the translation matrix $\boldsymbol{T}$. As shown below, this procedure produces reproducible, well-converged bounds across independent initializations.

To construct a numerical bound in the $\sigma^\star$--$\mu^\star$ plane, we decompose the diagonal matrix $y(\boldsymbol{D}^\star_\diamond)$ into contributions associated with the effective conductivity, bulk modulus, and shear modulus:
\begin{equation}
y(\boldsymbol{D}^\star_\diamond)
=
\boldsymbol{I}_\kappa z_\kappa
+
\boldsymbol{I}_\mu z_\mu
+
\boldsymbol{I}_\sigma z_\sigma.
\label{eq:y transform identity form}
\end{equation}
Depending on the bound $\diamond=\epsilon E,\epsilon J,\tau E,\tau J$, the parameters $z_\kappa,z_\mu,z_\sigma\in[0,\infty)$ are set to $y(\kappa^\star)$ or $1/y(\kappa^\star)$, $y(\mu^\star)$ or $1/y(\mu^\star)$, and $y(\sigma^\star)$ or $1/y(\sigma^\star)$, respectively, and are introduced for further notational convenience. The matrices $\boldsymbol{I}_\kappa$, $\boldsymbol{I}_\mu$, and $\boldsymbol{I}_\sigma$ are constant diagonal selector matrices that place each transformed scalar into the appropriate entries of $y(\boldsymbol{D}^\star_\diamond)$. For a fixed choice of $\diamond$, they are determined directly from the diagonal structure of $\boldsymbol{D}^\star_\diamond$ and may be written as
\begin{equation}
\boldsymbol{I}_\kappa
=
\frac{\partial y(\boldsymbol{D}^\star_\diamond)}{\partial z_\kappa},
\qquad
\boldsymbol{I}_\mu
=
\frac{\partial y(\boldsymbol{D}^\star_\diamond)}{\partial z_\mu},
\qquad
\boldsymbol{I}_\sigma
=
\frac{\partial y(\boldsymbol{D}^\star_\diamond)}{\partial z_\sigma}.
\label{eq:y_selector_matrices}
\end{equation}

To trace the bound systematically, we optimize its position along a prescribed set of directions in the $(z_\sigma,z_\mu)$ plane. Along each direction, we locate the point at which the matrix inequality defining the bound (Eq.\eqref{eq:translation_bound_main_clean}) becomes active and optimize the translation matrix $\boldsymbol{T}$ to move this point as far outward as permitted by the admissibility conditions. Under the inverse Y-transform, this corresponds to making the bound in the physical $(\sigma^\star,\mu^\star)$ plane as tight as possible along the corresponding curve. Repeating this procedure over a range of directions traces the boundary of the admissible effective-property region. Among the objective formulations we considered, this ray-based construction provided the most stable and consistently tight bounds. 

Accordingly, we parameterize the $(z_\sigma,z_\mu)$ plane by rays
\[
z_\sigma=\alpha z_\mu,
\qquad \alpha>0.
\]
For fixed $\boldsymbol{T}$ and fixed slope $\alpha$, let $z_\mu^o(\boldsymbol{T}\mid\alpha)$ denote the unique value at which the smallest eigenvalue reaches zero:
\begin{equation}
\lambda_{\min}\!\Big(
z_\kappa \boldsymbol{I}_\kappa
+
\big(\boldsymbol{I}_\mu+\alpha \boldsymbol{I}_\sigma\big) z_\mu^o(\boldsymbol{T}\mid\alpha)
+
\boldsymbol{T}
\Big)=0,
\label{eq:zmu_root_def}
\end{equation}
which denotes the point along the ray at which Eq.\eqref{eq:translation_bound_main_clean} becomes active. The corresponding point
\[
\big(\alpha z_\mu^o(\boldsymbol{T}\mid\alpha),\,z_\mu^o(\boldsymbol{T}\mid\alpha)\big)
\]
therefore gives the location of the bound along that ray.

The tightest bound on a given ray is obtained by maximizing $z_\mu^o(\boldsymbol{T}\mid\alpha)$. Indeed, $z_\mu^o$ is defined by the point at which the minimum eigenvalue vanishes, so making the matrix inside Eq.~\eqref{eq:zmu_root_def} smaller in the Loewner sense pushes the zero crossing to larger $z_\mu$. In particular, if $\boldsymbol{T}'\preceq \boldsymbol{T}$, then for fixed $(z_\kappa,z_\mu,\alpha)$,
\[
\lambda_{\min}\!\Big(
z_\kappa \boldsymbol{I}_\kappa
+
(\boldsymbol{I}_\mu+\alpha \boldsymbol{I}_\sigma) z_\mu
+
\boldsymbol{T}'
\Big)
\le
\lambda_{\min}\!\Big(
z_\kappa \boldsymbol{I}_\kappa
+
(\boldsymbol{I}_\mu+\alpha \boldsymbol{I}_\sigma) z_\mu
+
\boldsymbol{T}
\Big),
\]
and thus a larger value of $z_\mu$ is required to recover $\lambda_{\min}=0$. Thus, whenever two admissible translation matrices
are ordered in the Loewner sense, the smaller
matrix produces an equally strong or stronger
bound since this choice pushes the boundary farthest outward along the ray.

The remaining issue is the dependence on $z_\kappa$. Because our goal is a bound in the $\sigma^\star$--$\mu^\star$ plane, the final result must not depend on the unknown value of $\kappa^\star$. We therefore choose $z_\kappa$ so as to produce the \emph{loosest} admissible bound, ensuring that the bound is not artificially tightened by a particular choice of effective bulk modulus.

To see which choice is correct, consider the matrix
\[
\boldsymbol{M}(z_\kappa,z_\mu)
=
z_\kappa \boldsymbol{I}_\kappa
+
\big(\boldsymbol{I}_\mu+\alpha\boldsymbol{I}_\sigma\big) z_\mu
+
\boldsymbol{T}.
\]
Since both $\boldsymbol{I}_\kappa$ and $\boldsymbol{I}_\mu+\alpha\boldsymbol{I}_\sigma$ are positive semidefinite, $\lambda_{\min}(\boldsymbol{M})$ is nondecreasing in both $z_\kappa$ and $z_\mu$. It follows that, for fixed $\boldsymbol{T}$ and $\alpha$, increasing $z_\kappa$ raises the minimum eigenvalue and therefore decreases the root $z_\mu^o(\boldsymbol{T}\mid\alpha)$ defined by Eq.~\eqref{eq:zmu_root_def}. In other words, increasing $z_\kappa$ moves the bound inward and makes it looser. Therefore the loosest possible $\sigma^\star$--$\mu^\star$ bound is obtained in the limit
\[
z_\kappa\to\infty.
\]

Equivalently, sending $z_\kappa$ to infinity penalizes the bulk-modulus sector so strongly that it no longer influences the minimum-eigenvalue condition defining the $\sigma^\star$--$\mu^\star$ boundary. The resulting bound is therefore independent of $\kappa^\star$ and cannot be spuriously tightened by any finite choice of bulk modulus. In practice, we implement this by taking $z_\kappa$ to be sufficiently large that the computed bound is numerically unchanged upon further increase.

Although the bound could be optimized independently for each slope $\alpha$, doing so would require a separate optimization for every point along the boundary and would therefore be computationally expensive. Instead, we select a small set of slopes $\{\alpha\}$ and optimize a single objective that sums the bound positions along the corresponding rays:
\begin{equation}
\max_{\boldsymbol{T}\in\mathcal{S}_{\boldsymbol{T}}}\ \ \sum_{\alpha} z_\mu^o(\boldsymbol{T}\mid \alpha),
\qquad
\text{where } z_\mu^o(\boldsymbol{T}\mid \alpha)\ \text{satisfies Eq.~\eqref{eq:zmu_root_def} for each }\alpha.
\label{eq:objective_final_clean}
\end{equation}
Using a small number of well-separated slopes provides a favorable balance between computational cost and accuracy. In practice, we find that three to four slopes are generally sufficient to recover the full bound to the numerical resolution reported here. We optimize Eq.~\eqref{eq:objective_final_clean} over the feasible set
\[
\mathcal{S}_{\boldsymbol{T}}
=
\left\{
\boldsymbol{T}:
\ \boldsymbol{D}^i_{\diamond}-\boldsymbol{T}\succeq 0\ (i=1,2),
\ \boldsymbol{S}_\diamond^T\boldsymbol{U}(\underline{R})^T
\boldsymbol{T}\boldsymbol{U}(\underline{R})
\boldsymbol{S}_\diamond \succeq 0\ \forall \underline{R}\in\mathrm{SO}(3)
\right\}
\]
using a projected quasi-Newton method~\cite{schmidt2009}, with semidefinite feasibility checks performed in CVXPY~\cite{diamond2016cvxpy,agrawal2018rewriting}. We warn that using multiple slopes in the objective Eq.~\eqref{eq:objective_final_clean} can result in suboptimal bounds. If one desires the tightest possible bounds, we recommend using only one slope in the objective. 

As noted before, two aspects of the numerical procedure require explicit convergence checks. First, because the optimization over $\boldsymbol{T}$ is not guaranteed \emph{a priori} to reach the global optimum from a single initialization, we test whether independent random initializations converge to the same bound. Second, because the quasiconvexity condition in Eq.~\eqref{eq:quasi-convexity psd form} is formally indexed by infinitely many rotations, we verify that the finite rotation sampling introduced above is sufficiently dense.

Here, we provide both of these checks. We examine the numerical convergence of the bound point obtained along a fixed ray in the \((z_\sigma,z_\mu)\) plane. We consider the case with \(\kappa_1=\mu_1=\sigma_2=1.0\) and \(\kappa_2=\mu_2=\sigma_1=0.1\), and optimize the bound along the ray \(z_\sigma=\alpha z_\mu\) with \(\alpha=1\). For this choice, the inverse relation defined by the \(Y\)-transform maps the line \(z_\sigma=z_\mu\) to the line \(\sigma^\star=\mu^\star\) in physical property space. Thus, along this ray, the optimized values of \(\sigma^\star\) and \(\mu^\star\) coincide.

\begin{figure}[tb]
    \centering
    \includegraphics[width=0.99\linewidth]{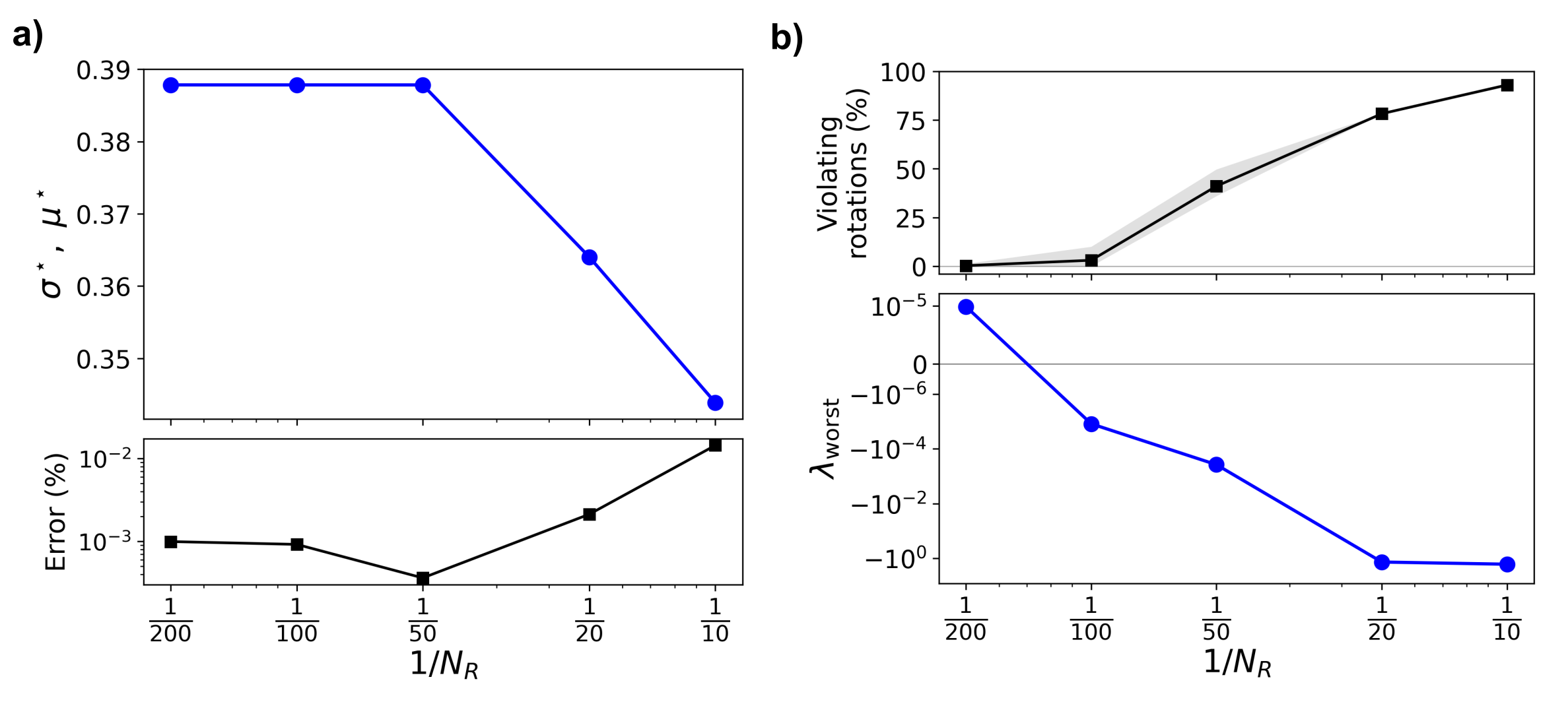}
    \caption{Convergence with the number $N_R$ of rotation constraints for the badly ordered case $\kappa_1=\mu_1=\sigma_2=1.0$, $\kappa_2=\mu_2=\sigma_1=0.1$, at $f=0.5$ and along $z_\sigma=z_\mu$. (a) Mean physical bound point $\sigma^\star=\mu^\star$ (top) and its relative sample standard deviation across independent optimizations (bottom). (b) Percentage of sampled rotations with $\lambda_{\min}(\boldsymbol{S}_\diamond^T\boldsymbol{U}(\underline{R})^T
\boldsymbol{T}\boldsymbol{U}(\underline{R})
\boldsymbol{S}_\diamond)<-10^{-8}$ (top; shading spans the range over independent optimizations), and the mean smallest eigenvalue found per translation matrix, $\lambda_{\mathrm{worst}}$ (bottom).}
    \label{fig:convergence}
\end{figure}

We assess both the dependence on initialization and the finite sampling of rotations in Eq.~\eqref{eq:quasi-convexity psd form}. For each $N_R$, we perform independent optimizations with an approximately $10^{-3}$ projected-gradient tolerance. Figure~\ref{fig:convergence}(a) shows that the physical bound point changes negligibly for $N_R\gtrsim50$, with relative variation across runs of order $10^{-3}\%$. Here the plotted ``Error'' is the sample standard deviation divided by the mean, expressed as a percentage.

Figure~\ref{fig:convergence}(b) tests quasiconvexity violation using $10^5$ independent Haar-distributed rotations per Translation matrix, and local minimization from 24 low-eigenvalue starting points. We note that the quasiconvexity constraint is satisfied when the smallest eigenvalue of $\boldsymbol{S}_\diamond^T\boldsymbol{U}(\underline{R})^T
\boldsymbol{T}\boldsymbol{U}(\underline{R})
\boldsymbol{S}_\diamond$ is nonnegative for all rotations. We denote the most violating eigenvalue found over all rotations by $\lambda_{\mathrm{worst}}=\lambda_{\min}(\boldsymbol{S}_\diamond^T\boldsymbol{U}(\underline{R}_*)^T
\boldsymbol{T}\boldsymbol{U}(\underline{R}_*)
\boldsymbol{S}_\diamond)$, where $\underline{R}_*=\arg\min_{\underline{R}\in\mathrm{SO(3)}}\lambda_{\min}(\boldsymbol{S}_\diamond^T\boldsymbol{U}(\underline{R})^T
\boldsymbol{T}\boldsymbol{U}(\underline{R})
\boldsymbol{S}_\diamond)$ is the rotation matrix which yields the smallest eigenvalue. The upper plot averages the percentage of violating rotations across optimization runs, while the lower plot averages $\lambda_{\mathrm{worst}}$ across those runs. From $N_R=50$ to $200$, the average smallest eigenvalue found across runs improves from a small negative value $-10^{-4}$ to a positive value $10^{-5}$. These tests provide strong numerical
evidence for convergence of the quasiconvexity
constraints, with the computed bound remaining
essentially unchanged for $N_R\gtrsim50$.

\section{Bound invariance under rotations of $\boldsymbol{T}$}
\label{app:bound_invce_under_rot_of_T}

Since the microscopic phases and macroscopic composite are isotropic,
the phase matrices $\boldsymbol{D}_\diamond^i$ for $i=1,2$ and $\boldsymbol{D}_\diamond^\star$ satisfy
\begin{equation}
    \boldsymbol{U}(\underline{R})^T
    \boldsymbol{D}
    \boldsymbol{U}(\underline{R})
    =\boldsymbol{D},
    \label{eq:D_invariance}
\end{equation}
where we have suppressed the $\star,i,\diamond$ for brevity. We note that $y(\boldsymbol{D})$ has the same property since the function $y(\cdot)$ acts component-wise.
To see this, it is convenient to express $\boldsymbol{D}$ in the general form:
\begin{equation}
    \boldsymbol{D} =
    \left[\boldsymbol{I}_5 \otimes
    \left(d_{\mathrm{hydro}} \oplus
    d_{\mathrm{dev}}\boldsymbol{I}_5\right)\right]
    \oplus d_{\mathrm{elec}}\boldsymbol{I}_9.
    \label{eq:D_gen_form}
\end{equation}
where $d_{\mathrm{hydro}}$, $d_{\mathrm{dev}}$, and $d_{\mathrm{elec}}$ are constants associated with the bulk modulus, shear modulus, and conductivity, respectively. Similarly, we may express the transformation $\boldsymbol{U}(\underline{R})$ as
\begin{equation}
    \boldsymbol{U}(\underline{R}) = [\boldsymbol{G}_{\mathrm{dev}}(\underline{R})\otimes(1\oplus\boldsymbol{G}_{\mathrm{dev}}(\underline{R}))]\oplus\boldsymbol{G}(\underline{R}).
\end{equation}
Eq.\eqref{eq:D_invariance} thus follows from direct product and direct sum algebra. 

Recall the translation method bound Eq.~\eqref{eq:translation_bound_main_clean} and the phase constraints Eq.~\eqref{eq:phase_constraints_clean}. Since $\boldsymbol{U}(\underline{R})$ is orthogonal, we preserve the inequalities if we multiply on the left by $\boldsymbol{U}(\underline{R})^T$ and on the right by $\boldsymbol{U}(\underline{R})$, giving
\begin{align}
    y(\boldsymbol{D}_{\diamond}^\star)+\boldsymbol{U}(\underline{R})^T\boldsymbol{T}\boldsymbol{U}(\underline{R})&\succeq0 \\
    \boldsymbol{D}_{\diamond}^i - \boldsymbol{U}(\underline{R})^T\boldsymbol{T}\boldsymbol{U}(\underline{R})&\succeq0,\quad i=1,2.
    \label{eq:U-transformed bounds no symmetry}
\end{align}
Now let us consider the quasiconvexity condition Eq.~\eqref{eq:quasi-convexity psd form}. Since $\boldsymbol{U}(\underline{R})$ is a representation of $\mathrm{SO(3)}$, for two rotation matrices $\underline{R}$ and $\underline{R}'$, we have $\boldsymbol{U}(\underline{R} \ \underline{R}')=\boldsymbol{U}(\underline{R})\boldsymbol{U}(\underline{R}')$. It thus follows that if $\boldsymbol{T}$ satisfies the quasiconvexity constraint, then so will its rotated counterpart $\boldsymbol{U}(\underline{R})^T\boldsymbol{T}\boldsymbol{U}(\underline{R})$. Thus, if a translation matrix $\boldsymbol{T}$ satisfies the macroscopic bound, phase constraints, and quasiconvexity condition, then its rotated counterpart $\boldsymbol{U}(\underline{R})^T\boldsymbol{T}\boldsymbol{U}(\underline{R})$ leaves the bound unchanged and remains feasible.

\section{Representation theory}
\label{app:representation_theory}

Recall that the rotation matrix $\boldsymbol{G}(\underline{R})$ can be decomposed as
\begin{align}
    \boldsymbol{G}(\underline{R})
    &=\boldsymbol{G}_{\mathrm{sym}}(\underline{R})
    \oplus\boldsymbol{G}_{\mathrm{asym}}(\underline{R})
    \notag\\
    &=1\oplus\boldsymbol{G}_{\mathrm{dev}}(\underline{R})
    \oplus\boldsymbol{G}_{\mathrm{asym}}(\underline{R}).
\end{align}
Consequently, after a suitable reordering of basis vectors, the $39$-dimensional representation $\boldsymbol{U}(\underline{R})$ takes the block-diagonal form
\begin{equation}
    \boldsymbol{U}'(\underline{R})=
    \boldsymbol{G}_{\mathrm{dev}}(\underline{R})
    \oplus
    \left(\boldsymbol{G}_{\mathrm{dev}}(\underline{R})
    \otimes\boldsymbol{G}_{\mathrm{dev}}(\underline{R})\right)
    \oplus 1
    \oplus\boldsymbol{G}_{\mathrm{dev}}(\underline{R})
    \oplus\boldsymbol{G}_{\mathrm{asym}}(\underline{R}),
    \label{eq:U_decomp_prime_permuted}
\end{equation}
where we use a prime to indicate that we have permuted some of the rows and columns of $\boldsymbol{U}(\underline{R})$.
These five blocks are invariant under rotations, meaning that vectors belonging to one block remain within that block under the action of $\boldsymbol{U}'(\underline{R})$.

To identify the irreducible components of this representation, recall that the irreducible representations of $\mathrm{SO(3)}$ are labeled by nonnegative integers $\ell=0,1,2,\ldots$, with each representation $\mathcal{V}_\ell$ having dimension $2\ell+1$. In particular, the scalar representation corresponds to $\mathcal{V}_0$, the three-dimensional antisymmetric tensor representation to $\mathcal{V}_1$, and the five-dimensional symmetric-deviatoric tensor representation to $\mathcal{V}_2$. These representations are already irreducible. Therefore, the only block that requires further decomposition is
$\boldsymbol{G}_{\mathrm{dev}}(\underline{R})\otimes\boldsymbol{G}_{\mathrm{dev}}(\underline{R})$.

A standard result of $\mathrm{SO(3)}$ representation theory, known as the angular-momentum addition rule, states that
\begin{equation}
    \mathcal{V}_{\ell_1}\otimes\mathcal{V}_{\ell_2}
    =\bigoplus_{\ell=|\ell_1-\ell_2|}^{\ell_1+\ell_2}
    \mathcal{V}_\ell.
\end{equation}
Applying this rule to two copies of $\mathcal{V}_2$ gives
\begin{equation}
    \mathcal{V}_2\otimes\mathcal{V}_2
    =\mathcal{V}_0\oplus\mathcal{V}_1
    \oplus\mathcal{V}_2\oplus\mathcal{V}_3
    \oplus\mathcal{V}_4.
\end{equation}
The dimensions of these five representations are $1,3,5,7,9$, respectively, which sum to $25$, as expected. We now construct an orthogonal transformation that separates these irreducible components.

We begin with the three standard generators of rotations,
\begin{equation}
    \underline{\Omega}_x=
    \begin{pmatrix}
        0&0&0\\
        0&0&-1\\
        0&1&0
    \end{pmatrix},
    \quad
    \underline{\Omega}_y=
    \begin{pmatrix}
        0&0&1\\
        0&0&0\\
        -1&0&0
    \end{pmatrix},
    \quad
    \underline{\Omega}_z=
    \begin{pmatrix}
        0&-1&0\\
        1&0&0\\
        0&0&0
    \end{pmatrix},
\end{equation}
which generate rotations according to
$\underline{R}_i(\theta)=e^{\theta\underline{\Omega}_i}$.
The corresponding generators of
$\boldsymbol{G}_{\mathrm{dev}}(\underline{R})$,
denoted by $\boldsymbol{g}_i$ for $i=x,y,z$, have components
\begin{align}
    g_i^{(\alpha\beta)}
    &=
    \left.
    \frac{d}{d\theta}
    G_{\mathrm{dev}}^{(\alpha\beta)}
    \left(e^{\theta\underline{\Omega}_i}\right)
    \right|_{\theta=0}
    \notag\\
    &=
    \underline{a}^{(\alpha+1)}:
    [\underline{\Omega}_i,\underline{a}^{(\beta+1)}],
    \qquad \alpha,\beta=1,\ldots,5,
\end{align}
where $[\underline{a},\underline{b}]
=\underline{a}\underline{b}-\underline{b}\underline{a}$
denotes the matrix commutator.

The generators of the tensor-product representation
$\boldsymbol{G}_{\mathrm{dev}}(\underline{R})
\otimes\boldsymbol{G}_{\mathrm{dev}}(\underline{R})$
are then given by
\begin{equation}
    \boldsymbol{\mathfrak{g}}_i
    =\boldsymbol{g}_i\otimes\boldsymbol{I}_5
    +\boldsymbol{I}_5\otimes\boldsymbol{g}_i.
\end{equation}
Using these generators, we construct the $25\times25$ Casimir operator
\begin{equation}
    \boldsymbol{\mathcal{C}}
    =-\boldsymbol{\mathfrak{g}}_x^2
    -\boldsymbol{\mathfrak{g}}_y^2
    -\boldsymbol{\mathfrak{g}}_z^2.
\end{equation}
The useful property of the Casimir operator is that it commutes with each generator,
\begin{equation}
    [\boldsymbol{\mathcal{C}},
    \boldsymbol{\mathfrak{g}}_i]=0,
    \qquad i=x,y,z.
\end{equation}
Consequently, its eigenspaces are invariant under rotations. Furthermore, the Casimir operator acts on each irreducible representation $\mathcal{V}_\ell$ as multiplication by $\ell(\ell+1)$. Since each representation appears exactly once in the decomposition of $\mathcal{V}_2\otimes\mathcal{V}_2$, the eigenspaces of $\boldsymbol{\mathcal{C}}$ correspond precisely to the five irreducible components. In particular, the eigenvalues of $\boldsymbol{\mathcal{C}}$ are
\begin{equation}
    \lambda_\ell=\ell(\ell+1),
    \qquad \ell=0,1,2,3,4,
\end{equation}
giving eigenvalues $0,2,6,12,20$ with multiplicities $1,3,5,7,9$, respectively. The corresponding orthonormal eigenvectors provide the desired orthogonal decomposition of the $25$-dimensional tensor-product representation.

Let $\boldsymbol{E}_\ell$ denote the matrix whose columns
are an orthonormal basis for the eigenspace of
$\boldsymbol{\mathcal{C}}$ with eigenvalue $\ell(\ell+1)$.
Thus, $\boldsymbol{E}_\ell$ has dimensions
$25\times(2\ell+1)$, and the matrix
\begin{equation}
    \boldsymbol{Q}_{25}^{(0)}
    =
    \begin{pmatrix}
        \boldsymbol{E}_0 &
        \boldsymbol{E}_1 &
        \boldsymbol{E}_2 &
        \boldsymbol{E}_3 &
        \boldsymbol{E}_4
    \end{pmatrix}
\end{equation}
is orthogonal. It transforms the tensor-product
representation into five irreducible blocks,
\begin{align}
    &(\boldsymbol{Q}_{25}^{(0)})^T
    \left(\boldsymbol{G}_{\mathrm{dev}}(\underline{R})
    \otimes\boldsymbol{G}_{\mathrm{dev}}(\underline{R})\right)
    \boldsymbol{Q}_{25}^{(0)}
    \notag\\
    &\qquad =
    1\oplus\boldsymbol{F}_1(\underline{R})
    \oplus\boldsymbol{F}_2(\underline{R})
    \oplus\boldsymbol{F}_3(\underline{R})
    \oplus\boldsymbol{F}_4(\underline{R}).
\end{align}
However, the bases within the Casimir eigenspaces
are not uniquely determined. In particular,
$\boldsymbol{F}_1(\underline{R})$ and $\boldsymbol{F}_2(\underline{R})$ are
equivalent representations to $\boldsymbol{G}_{\mathrm{asym}}(\underline{R})$
and $\boldsymbol{G}_{\mathrm{dev}}(\underline{R})$, respectively,
but are not generally expressed in the same bases. Since $\boldsymbol{G}_{\mathrm{asym}}(\underline{R})$
and $\boldsymbol{G}_{\mathrm{dev}}(\underline{R})$ appear in other blocks of $\boldsymbol{U}(\underline{R})$, it is therefore of interest to transform $\boldsymbol{F}_1(\underline{R})$ and $\boldsymbol{F}_2(\underline{R})$ into $\boldsymbol{G}_{\mathrm{asym}}(\underline{R})$ and $\boldsymbol{G}_{\mathrm{dev}}(\underline{R})$, respectively. To perform this alignment, we first construct the generators of $\boldsymbol{F}_1(\underline{R})$ and $\boldsymbol{F}_2(\underline{R})$ as
\begin{equation}
    \boldsymbol{h}_i^{\ell}
    =\boldsymbol{E}_\ell^T
    \boldsymbol{\mathfrak{g}}_i
    \boldsymbol{E}_\ell.
\end{equation}
We then determine orthogonal matrices
$\boldsymbol{W}_1$ and $\boldsymbol{W}_2$
satisfying
\begin{align}
    \boldsymbol{h}_i^{(1)}\boldsymbol{W}_1
    &=\boldsymbol{W}_1\boldsymbol{g}_i^{\mathrm{asym}},
    \\
    \boldsymbol{h}_i^{(2)}\boldsymbol{W}_2
    &=\boldsymbol{W}_2\boldsymbol{g}_i^{\mathrm{dev}},
\end{align}
for $i=x,y,z$, where
$\boldsymbol{g}_i^{\mathrm{asym}}$ and
$\boldsymbol{g}_i^{\mathrm{dev}}$ are the generators
of the corresponding original rotation matrices.
These equations are linear in the entries of
$\boldsymbol{W}_1$ and $\boldsymbol{W}_2$,
and their nonzero solutions can be normalized
to orthogonal matrices.

The aligned orthogonal transformation is therefore
\begin{equation}
    \boldsymbol{Q}_{25}
    =
    \begin{pmatrix}
        \boldsymbol{E}_0 &
        \boldsymbol{E}_1\boldsymbol{W}_1 &
        \boldsymbol{E}_2\boldsymbol{W}_2 &
        \boldsymbol{E}_3 &
        \boldsymbol{E}_4
    \end{pmatrix}.
\end{equation}
In this basis, the first three irreducible
blocks are represented by
$1$, $\boldsymbol{G}_{\mathrm{asym}}(\underline{R})$,
and $\boldsymbol{G}_{\mathrm{dev}}(\underline{R})$,
respectively. No further alignment is required
for the remaining two blocks because each
occurs only once in the full representation of $\boldsymbol{U}(\underline{R})$. We can thus assemble $\boldsymbol{Q}_{25}$ into an orthogonal transformation acting on the full 39-dimensional space,
\begin{equation}
    \boldsymbol{Q}_{39}=\boldsymbol{I}_5\oplus\boldsymbol{Q}_{25}\oplus\boldsymbol{I}_9.
\end{equation}
Finally, the Clebsch-Gordan transformation $\boldsymbol{Q}_{\mathrm{CG}}$ from the main text is obtained by permuting this matrix such that the sectors with common representation $\mathcal{V}_l$ associated with $\boldsymbol{U}(\underline{R})$ are grouped. The irreducible form of $\boldsymbol{U}_{\mathrm{CG}}(\underline{R})$ in Eq.~\eqref{eq:U_CG} thus follows.

We can now determine the form of the property matrices
in the Clebsch--Gordan basis. Under the basis permutation
used to obtain Eq.~\eqref{eq:U_decomp_prime_permuted},
the generic property matrix Eq.~\eqref{eq:D_gen_form}
becomes
\begin{equation}
    \boldsymbol{D}'
    =d_{\mathrm{hydro}}\boldsymbol{I}_5
    \oplus d_{\mathrm{dev}}\boldsymbol{I}_{25}
    \oplus d_{\mathrm{elec}}\boldsymbol{I}_9.
\end{equation}
Since
$\boldsymbol{Q}_{39}^T\boldsymbol{D}'\boldsymbol{Q}_{39}
=\boldsymbol{D}'$,
the transformation preserves the hydrostatic,
deviatoric, and electrical property sectors.
Grouping the equivalent irreducible representations
through the final basis permutation then gives
\begin{equation}
\begin{aligned}
\boldsymbol{D}_{\mathrm{CG}}
=
\begin{pmatrix}
d_{\mathrm{dev}} & 0\\
0 & d_{\mathrm{elec}}
\end{pmatrix}
\oplus
\left[
\begin{pmatrix}
d_{\mathrm{dev}} & 0\\
0 & d_{\mathrm{elec}}
\end{pmatrix}
\otimes\boldsymbol{I}_3
\right]
\oplus
\left[
\begin{pmatrix}
d_{\mathrm{hydro}} & 0 & 0\\
0 & d_{\mathrm{dev}} & 0\\
0 & 0 & d_{\mathrm{elec}}
\end{pmatrix}
\otimes\boldsymbol{I}_5
\right]
\oplus d_{\mathrm{dev}}\boldsymbol{I}_7
\oplus d_{\mathrm{dev}}\boldsymbol{I}_9.
\end{aligned}
\label{eq:D_CG}
\end{equation}
This form applies to both phase and effective property
matrices, with the corresponding scalar coefficients
$d_{\mathrm{hydro}}$, $d_{\mathrm{dev}}$, and
$d_{\mathrm{elec}}$.

\section{Finding tight bounds with symmetry reduction}
\label{app:sym_red_bounds}

Recall that the symmetry-reduced translation bound and its constraints are given by Eq.~\eqref{eq:CG_bounds}. To write these inequalities explicitly, we parameterize
the symmetric translation blocks as
\begin{equation}
\begin{aligned}
\boldsymbol{t}_0 &=
\begin{pmatrix}
t_{0,1} & t_{0,3}\\
t_{0,3} & t_{0,2}
\end{pmatrix},
\qquad
\boldsymbol{t}_1 =
\begin{pmatrix}
t_{1,1} & t_{1,3}\\
t_{1,3} & t_{1,2}
\end{pmatrix},\qquad
\boldsymbol{t}_2 =
\begin{pmatrix}
t_{2,1} & t_{2,4} & t_{2,5}\\
t_{2,4} & t_{2,2} & t_{2,6}\\
t_{2,5} & t_{2,6} & t_{2,3}
\end{pmatrix}.
\end{aligned}
\label{eq:t_components}
\end{equation}
Using the block-diagonal forms of
$y(\boldsymbol{D}^\star_\diamond)_{\mathrm{CG}}$
and $\boldsymbol{T}_{\mathrm{CG}}$, and suppressing $\diamond$ for brevity, the macroscopic
bound separates into the following five conditions:
\begin{subequations}
\label{eq:CG_macro_blocks}
\begin{align}
&\begin{pmatrix}
y(d^\star_{\mathrm{dev}})+t_{0,1} & t_{0,3}\\
t_{0,3} & y(d^\star_{\mathrm{elec}})+t_{0,2}
\end{pmatrix}
\succeq 0,
\label{eq:CG_macro_V0}
\\[8pt]
&\begin{pmatrix}
y(d^\star_{\mathrm{dev}})+t_{1,1} & t_{1,3}\\
t_{1,3} & y(d^\star_{\mathrm{elec}})+t_{1,2}
\end{pmatrix}
\succeq 0,
\label{eq:CG_macro_V1}
\\[8pt]
&\begin{pmatrix}
y(d^\star_{\mathrm{hydro}})+t_{2,1} & t_{2,4} & t_{2,5}\\
t_{2,4} & y(d^\star_{\mathrm{dev}})+t_{2,2} & t_{2,6}\\
t_{2,5} & t_{2,6} & y(d^\star_{\mathrm{elec}})+t_{2,3}
\end{pmatrix}
\succeq 0,
\label{eq:CG_macro_V2}
\\[8pt]
&y(d^\star_{\mathrm{dev}})+t_3\geq 0,
\label{eq:CG_macro_V3}
\\
&y(d^\star_{\mathrm{dev}})+t_4\geq 0.
\label{eq:CG_macro_V4}
\end{align}
\end{subequations}
These conditions correspond to the irreducible
representations $\mathcal{V}_0,\mathcal{V}_1,
\mathcal{V}_2,\mathcal{V}_3$, and $\mathcal{V}_4$,
respectively. Since we are only interested in the $\sigma^\star$--$\mu^\star$ bounds, we set $y(d^\star_{\mathrm{hydro}})$ equal to infinity analogous to the procedure in \ref{app:old_numerical_approach}. This converts the bound Eq.~\eqref{eq:CG_macro_V2} into
\begin{equation}
    \begin{pmatrix}
    y(d^\star_{\mathrm{dev}})+t_{2,2} & t_{2,6}\\
    t_{2,6} & y(d^\star_{\mathrm{elec}})+t_{2,3}
    \end{pmatrix}\succeq0.
\end{equation}

Similarly, the phase admissibility conditions
$\boldsymbol{D}^i_{\mathrm{CG},\diamond}
-\boldsymbol{T}_{\mathrm{CG}}\succeq0$
reduce to
\begin{subequations}
\label{eq:CG_phase_blocks}
\begin{align}
&\begin{pmatrix}
d^i_{\mathrm{dev}}-t_{0,1} & -t_{0,3}\\
-t_{0,3} & d^i_{\mathrm{elec}}-t_{0,2}
\end{pmatrix}
\succeq 0,
\label{eq:CG_phase_V0}
\\[8pt]
&\begin{pmatrix}
d^i_{\mathrm{dev}}-t_{1,1} & -t_{1,3}\\
-t_{1,3} & d^i_{\mathrm{elec}}-t_{1,2}
\end{pmatrix}
\succeq 0,
\label{eq:CG_phase_V1}
\\[8pt]
&\begin{pmatrix}
d^i_{\mathrm{hydro}}-t_{2,1} & -t_{2,4} & -t_{2,5}\\
-t_{2,4} & d^i_{\mathrm{dev}}-t_{2,2} & -t_{2,6}\\
-t_{2,5} & -t_{2,6} & d^i_{\mathrm{elec}}-t_{2,3}
\end{pmatrix}
\succeq 0,
\label{eq:CG_phase_V2}
\\[8pt]
&d^i_{\mathrm{dev}}-t_3\geq 0,
\label{eq:CG_phase_V3}
\\
&d^i_{\mathrm{dev}}-t_4\geq 0,
\label{eq:CG_phase_V4}
\end{align}
\end{subequations}
for each phase $i=1,2$. 
Here, the coefficients
$d_{\mathrm{hydro}}$, $d_{\mathrm{dev}}$,
and $d_{\mathrm{elec}}$ of the phase and
effective property matrices are given by
\begin{equation}
\begin{array}{c|ccc|ccc}
& \multicolumn{3}{c|}{\text{Phase }i}
& \multicolumn{3}{c}{\text{Effective}}\\
\diamond
& d^i_{\mathrm{hydro}}
& d^i_{\mathrm{dev}}
& d^i_{\mathrm{elec}}
& d^\star_{\mathrm{hydro}}
& d^\star_{\mathrm{dev}}
& d^\star_{\mathrm{elec}}\\
\hline
\epsilon E
& 3\kappa_i & 2\mu_i & \sigma_i
& 3\kappa^\star & 2\mu^\star & \sigma^\star\\
\epsilon J
& 3\kappa_i & 2\mu_i & 1/\sigma_i
& 3\kappa^\star & 2\mu^\star & 1/\sigma^\star\\
\tau E
& 1/(3\kappa_i) & 1/(2\mu_i) & \sigma_i
& 1/(3\kappa^\star) & 1/(2\mu^\star) & \sigma^\star\\
\tau J
& 1/(3\kappa_i) & 1/(2\mu_i) & 1/\sigma_i
& 1/(3\kappa^\star) & 1/(2\mu^\star) & 1/\sigma^\star
\end{array}
\label{eq:scalar_coefficients}
\end{equation}
where $i=1,2$ denotes the material phase. Reinstating $\diamond$, the quasiconvexity condition remains
\begin{equation}
    \boldsymbol{S}_\diamond^T
    \boldsymbol{Q}_{\mathrm{CG}}
    \boldsymbol{T}_{\mathrm{CG}}
    \boldsymbol{Q}_{\mathrm{CG}}^T
    \boldsymbol{S}_\diamond\succeq 0.
\end{equation}
Although this condition can be further decomposed
into smaller positive-semidefinite blocks, their
explicit expressions are lengthy and depend on the choice of $\boldsymbol{Q}_{\mathrm{CG}}$. We therefore
omit them here, and note that the condition is enforced directly in our numerical calculations.

\begin{figure}[tb]
    \centering
    \includegraphics[width=0.99\linewidth]{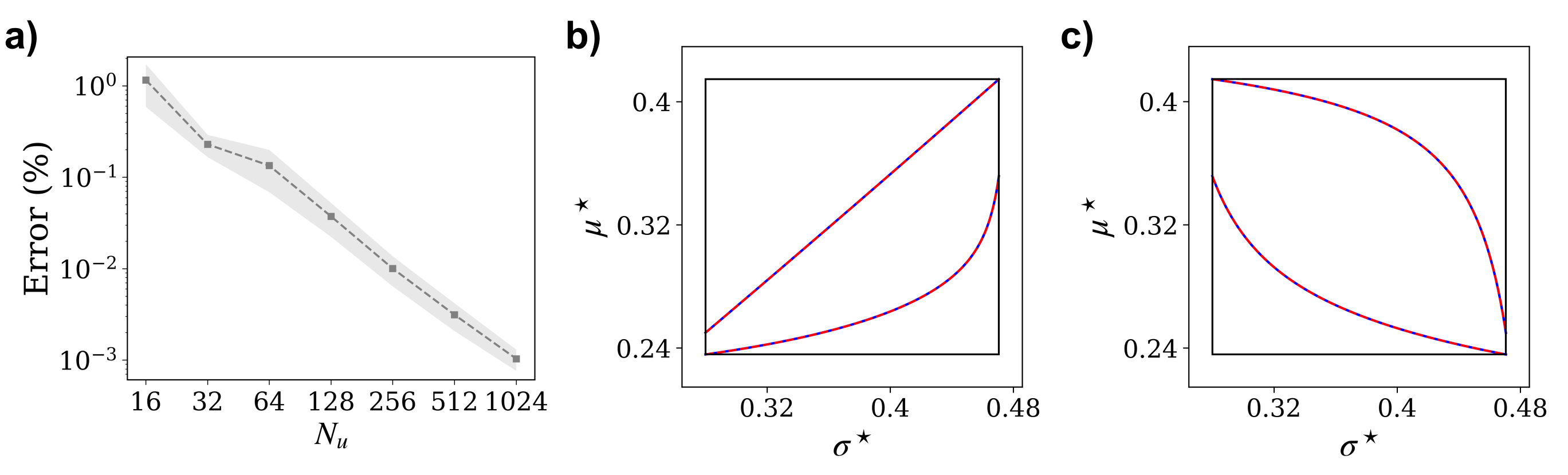}
    \caption{Numerical convergence and comparison of the bounds. (a) Convergence as the number of sampled directions $N_u$ increases. Each calculation is compared with a finer calculation using $N_u=2048$. The error is the largest difference in $\mu^\star$ along the boundary, expressed as a percentage of the interval between the lower and upper Hashin--Shtrikman shear bounds. We check four material settings, three volume fractions ($f=0.25,0.5,0.75$), and both sides of the boundary. The line shows the average error from eight randomized sampling grids, and the shaded band shows one standard deviation. Here $N_u$ sets the initial grid size; additional directions are included near the endpoints and during refinement. (b,c) Comparison of the symmetry-reduced bounds Eq.~\eqref{eq:intersection_of_u_feasible_region} (blue solid) with non-symmetry-reduced bounds Eq.~\eqref{eq:objective_final_clean} (red dashed). The symmetry-reduced calculations start with grids of 256 or 1024 directions and refine them further. Each single-ray optimization associated with the unrestricted bound starts from the corresponding symmetry-reduced solution. The red curves retain these starting bounds and include any stronger bounds found that pass additional quasiconvexity checks. The non-symmetry-reduced calculations initially enforce quasiconvexity at 50 sampled rotations, adding further rotations when violations are found. Black rectangles show the individual Hashin--Shtrikman bounds. Both panels use $f=0.5$, $\kappa_1=\mu_1=0.1$, and $\kappa_2=\mu_2=1$. The conductivities are $\sigma_1=0.1$, $\sigma_2=1$ in (b), and $\sigma_1=1$, $\sigma_2=0.1$ in (c).}
    \label{fig:convergence_sym_red}
\end{figure}

To determine the tightest bounds, we use the fact
that a symmetric matrix is positive semidefinite
if and only if its associated quadratic form is
nonnegative for every vector. Thus, the
macroscopic translation inequality is equivalent to
\begin{equation}
    \boldsymbol{u}^T
    y(\boldsymbol{D}^\star_\diamond)_{\mathrm{CG}}
    \boldsymbol{u}
    +
    \boldsymbol{u}^T
    \boldsymbol{T}_{\mathrm{CG}}
    \boldsymbol{u}
    \geq 0,
    \quad \text{for all }\,\boldsymbol{u}.
    \label{eq:u_on_ybound}
\end{equation}
Let $\boldsymbol{t}$ denote the vector of the 14
independent translation parameters, and let
$\mathcal{S}_{\boldsymbol{t}}$ be the set of
parameters satisfying all phase and quasiconvexity
constraints. For a fixed vector $\boldsymbol{u}$,
the strongest inequality is obtained by minimizing
the translation quadratic form:
\begin{equation}
    \Gamma(\boldsymbol{u})=
    -\min_{\boldsymbol{t}\in\mathcal{S}_{\boldsymbol{t}}}
    \boldsymbol{u}^T
    \boldsymbol{T}_{\mathrm{CG}}(\boldsymbol{t})
    \boldsymbol{u}.
    \label{eq:gamma_full}
\end{equation}
Because the objective is linear in the translation
parameters and the constraints are positive-semidefinite
matrix inequalities, this is a convex semidefinite
optimization problem. The optimized translation bound therefore requires
\begin{equation}
    \boldsymbol{u}^T
    y(\boldsymbol{D}^\star_\diamond)_{\mathrm{CG}}
    \boldsymbol{u}
    -\Gamma(\boldsymbol{u})\geq0,
    \quad \text{for all }\,\boldsymbol{u}.
    \label{eq:optimized_u_bound}
\end{equation}
Equivalently, the outer bound on the effective
properties is
\begin{equation}
    \mathcal{A}_\diamond=
    \bigcap_{\boldsymbol{u}}
    \Big\{(\mu^\star,\sigma^\star):
    \ \boldsymbol{u}^T
    y(\boldsymbol{D}^\star_\diamond)_{\mathrm{CG}}
    \boldsymbol{u}
    -\Gamma(\boldsymbol{u})\geq0
    \Big\}.
\label{eq:intersection_of_u_feasible_region}
\end{equation}
Since both the transformed property matrix and the
translation matrix are block diagonal, it is
sufficient to consider vectors within each
irreducible block separately. Furthermore, each
block has the form of a smaller matrix tensored
with an identity matrix. We therefore need only
consider vectors of dimensions $2$, $2$, $3$,
$1$, and $1$ for the five irreducible sectors,
respectively. In practice, we sample unit vectors within these
reduced spaces and solve Eq.~\eqref{eq:gamma_full}
for each direction. The intersection of the
resulting inequalities provides a numerical
approximation to the complete optimized
translation bound. 

In Figure~\ref{fig:convergence_sym_red}(a), we show that increasing the number of sampled directions $N_u$ improves the accuracy of the bound. We plot the mean error and a one-standard-deviation band relative to a finer calculation with $N_u=2048$. The error decreases rapidly, with $N_u\sim100$ sufficient to achieve below $0.1\%$ error. Figures~\ref{fig:convergence_sym_red}(b,c) show close agreement between the symmetry-reduced bounds in Eq.~\eqref{eq:intersection_of_u_feasible_region} and the non-symmetry-reduced bounds obtained using single-ray optimizations based on Eq.~\eqref{eq:objective_final_clean}.

To assess which irreducible sector determines the bound, we independently optimized the bounds associated with $\mathcal V_0,\ldots,\mathcal V_4$ for 600 material settings spanning different phase properties, contrasts, and orderings. All fourteen translation parameters remained free, and the full phase and quasiconvexity constraints were retained.  In every setting, the $\mathcal V_0$ bound was at least as restrictive as the remaining sectors within a relative numerical tolerance of $2\times10^{-5}$. These results provide extensive numerical evidence that the $\mathcal V_0$ inequality alone determines the symmetry-reduced bound.

\section{Derivation of parametric bounds}
\label{app:derivation_of_parametric_bounds}

Here we suppress the index $\diamond$ for brevity. We first note that the admissible translation set is invariant under reversing the signs of the electrical field components. Such a transformation reverses the elastic--electrical off-diagonal entries of $\boldsymbol{t}_0$ while preserving all phase and quasiconvexity constraints. Consequently, when optimizing the bound, it is sufficient to consider unit vectors with nonnegative components, parameterized by $\boldsymbol{u}=(\sqrt{\theta},\sqrt{1-\theta})^T$ for $\theta\in[0,1]$.

Taking the quadratic form of Eq.~\eqref{eq:reduced_CG_bound} with respect to $\boldsymbol{u}$ and optimizing over the admissible translation matrices gives
\begin{equation}
    y(d^\star_{\mathrm{dev}})\theta
    +y(d^\star_{\mathrm{elec}})(1-\theta)
    \geq\gamma(\theta),
    \qquad \forall\,\theta\in[0,1],
    \label{eq:final_inequality_theta}
\end{equation}
where $\gamma(\theta)$ is defined by Eq.~\eqref{eq:gamma_main_txt}. For each $\theta$, this inequality defines a half-plane in the transformed property space. The translation-method bound is therefore given by the intersection of these half-planes. The boundary can be constructed from the envelope of the corresponding lines,
\begin{equation}
    y(d^\star_{\mathrm{dev}})\theta
    +y(d^\star_{\mathrm{elec}})(1-\theta)
    =\gamma(\theta).
    \label{eq:final_equality_theta}
\end{equation}
At an interior point of the envelope, both this equation and its derivative with respect to $\theta$ must hold, with the transformed effective properties held fixed during differentiation. We therefore obtain
\begin{equation}
    y(d^\star_{\mathrm{dev}})
    -y(d^\star_{\mathrm{elec}})
    =\gamma'(\theta).
\end{equation}
Solving these two equations gives
\begin{align}
    y(d^\star_{\mathrm{dev}})
    &=\gamma(\theta)+(1-\theta)\gamma'(\theta),
    \notag\\
    y(d^\star_{\mathrm{elec}})
    &=\gamma(\theta)-\theta\gamma'(\theta).
\end{align}
This holds since, as demonstrated in \ref{app:gamma_differentiability}, $\gamma(\theta)$ is concave on $\theta\in[0,1]$ and differentiable throughout $\theta\in(0,1)$. 

Let $\boldsymbol{t}_0^\star(\theta)$ denote an optimal translation block in Eq.~\eqref{eq:gamma_main_txt}. By the envelope theorem, the derivative $\gamma'(\theta)$ can be evaluated by differentiating the objective while holding the optimal translation parameters fixed. This gives
\begin{equation}
\gamma'(\theta)
=-t_{0,1}^\star(\theta)
+t_{0,2}^\star(\theta)
-\frac{1-2\theta}{\sqrt{\theta(1-\theta)}}
t_{0,3}^\star(\theta).
\end{equation}
Substituting into the envelope equations yields the parametric expressions in Eq.~\eqref{eq:parametric_bound}. At the endpoints, Eq.~\eqref{eq:final_inequality_theta} gives
\begin{equation}
    y(d^\star_{\mathrm{elec}})\geq\gamma(0),
    \qquad
    y(d^\star_{\mathrm{dev}})\geq\gamma(1).
\end{equation}
The equalities define the corresponding limiting boundary lines in Eq.~\eqref{eq:parametric_bound_endpoints}.

\section{Concavity and differentiability of $\gamma(\theta)$}
\label{app:gamma_differentiability}

Here we suppress the bound classification index $\diamond$ for brevity. We show that $\gamma(\theta)$ defined by Eq.~\eqref{eq:gamma_main_txt} is concave on $[0,1]$ and differentiable for $0<\theta<1$. The key property is that feasible translations can be combined after separately scaling their mechanical and electrical fields. Specifically, let $\boldsymbol T^{(1)}$ and $\boldsymbol T^{(2)}$ be feasible and define
\begin{equation}
\overline{\boldsymbol T}=\sum_{j=1}^2\boldsymbol C_j\boldsymbol T^{(j)}\boldsymbol C_j,
\qquad
\boldsymbol C_j=\operatorname{diag}(\alpha_j\boldsymbol I_{30},\beta_j\boldsymbol I_9),
\qquad
\sum_j\alpha_j^2=\sum_j\beta_j^2=1.
\end{equation}
This combination remains feasible. Each $\boldsymbol C_j$ commutes with the phase matrices, giving
\[
\boldsymbol D^i_\diamond-\overline{\boldsymbol T}=\sum_j\boldsymbol C_j(\boldsymbol D^i_\diamond-\boldsymbol T^{(j)})\boldsymbol C_j\succeq0.
\]
Moreover, $\boldsymbol C_j\boldsymbol S_\diamond=\boldsymbol S_\diamond\widehat{\boldsymbol C}_j$, where $\widehat{\boldsymbol C}_j$ applies the same scalings to the independent mechanical and electrical amplitudes. Consequently,
\[
\boldsymbol S_\diamond^T\overline{\boldsymbol T}\boldsymbol S_\diamond
=\sum_j\widehat{\boldsymbol C}_j^T(\boldsymbol S_\diamond^T\boldsymbol T^{(j)}\boldsymbol S_\diamond)\widehat{\boldsymbol C}_j\succeq0.
\]
Finally, the scalings commute with $\boldsymbol U(\underline R)$, so rotational invariance is preserved and the reference quasiconvexity condition continues to hold for every rotation.

To establish concavity, choose optimizing translations at $\theta_1,\theta_2$, and set $\theta=\lambda\theta_1+(1-\lambda)\theta_2$, with $0<\theta<1$. Writing $\lambda_1=\lambda$ and $\lambda_2=1-\lambda$, choose the nonnegative scaling factors
\[
\alpha_j=\sqrt{\frac{\lambda_j\theta_j}{\theta}},
\qquad
\beta_j=\sqrt{\frac{\lambda_j(1-\theta_j)}{1-\theta}}.
\]
Their squared sums equal one. In the $V_0$ block, the diagonal entries scale by $\alpha_j^2,\beta_j^2$ and the off-diagonal entry by $\alpha_j\beta_j$. Substituting into the objective shows that $\overline{\boldsymbol T}$ achieves exactly $\lambda\gamma(\theta_1)+(1-\lambda)\gamma(\theta_2)$. Maximizing over all feasible translations therefore gives
\begin{equation}
\gamma(\lambda\theta_1+(1-\lambda)\theta_2)\geq\lambda\gamma(\theta_1)+(1-\lambda)\gamma(\theta_2).
\end{equation}
This proves concavity; mixtures equal to an endpoint satisfy the same inequality trivially.

Now choose any $\theta_0\in(0,1)$ and hold an optimizing translation there fixed. Its objective defines the smooth function
\[
f(\theta)=-\theta t_{0,1}^{*}(\theta_0)-(1-\theta)t_{0,2}^{*}(\theta_0)-2\sqrt{\theta(1-\theta)}t_{0,3}^{*}(\theta_0).
\]
By construction, $f(\theta)\leq\gamma(\theta)$ everywhere,
with equality at $\theta_0$. Thus, a smooth function
touches $\gamma$ from below at every interior point. Since $\gamma$ is concave, its left and right derivatives
exist and satisfy
\begin{equation}
\gamma'_-(\theta_0)\geq\gamma'_+(\theta_0).
\end{equation}
On the other hand, because $f(\theta)\leq\gamma(\theta)$
with equality at $\theta_0$, we have
\begin{equation}
\gamma'_-(\theta_0)
\leq f'(\theta_0)
\leq\gamma'_+(\theta_0).
\end{equation}
Combining these inequalities gives
\begin{equation}
\gamma'_-(\theta_0)
=\gamma'_+(\theta_0)
=f'(\theta_0).
\end{equation}
Therefore, $\gamma$ is differentiable for
$\theta\in(0,1)$.

\section{Topology optimization for isotropic composites}
\label{app:topopt}

Here we provide a self-contained description of the unit-cell topology optimization procedure we use to search for microstructures that approach the numerical cross-property bounds. Each optimized periodic cell is interpreted as an anisotropic grain within an aggregate of randomly oriented grains. The aggregate is therefore macroscopically isotropic, and its effective conductivity and elastic moduli are estimated using the self-consistent effective-medium approximation (EMA) \cite{markov2000,han2021}. This construction allows arbitrary anisotropic unit cells to be compared directly with the isotropic bounds.

We consider a periodic unit cell of volume $V=1$ containing a two-phase composite with phase properties $(\sigma_i,\kappa_i,\mu_i)$ for $i=1,2$. As before, $\rho(\mathbf r)\in\{0,1\}$ denotes the spatially dependent indicator function for phase 1 within the unit cell, and each local material property $x(r)=\rho(r)x_1+(1-\rho(r))x_2$. Although the physical design variable is binary, topology optimization is performed using the continuous relaxation $\rho(\mathbf r)\in[0,1]$. Intermediate values represent mixtures of the two phases during optimization, while the Solid Isotropic Material with Penalization (SIMP) method described below promotes convergence toward a binary design.

For conductivity, we minimize the electrostatic energy with respect to a periodic scalar potential $\phi$ under a prescribed constant macroscopic electric field $e_0$. The local electric and current fields are
\[
e=e_0-\nabla\phi,
\qquad
j=\sigma e,
\]
and the electrostatic energy functional is
\begin{equation}
H_{\mathrm{elec}}[\phi]
=\int j\cdot e\,dr,
\label{eq:electrostatic energy}
\end{equation}
where the integration domain is the periodic unit cell. The effective conductivity tensor $\underline{\Sigma}$ of the unit cell is defined by
\begin{equation}
\langle j\rangle=\underline{\Sigma}\,e_0,
\label{eq:avg_current_conductivity}
\end{equation}
where $\langle\cdot\rangle$ denotes a volume average over the cell. Because $\underline{\Sigma}\in\mathbb{R}^{3\times3}$, we solve Eq.~\eqref{eq:electrostatic energy} for three mutually orthogonal macroscopic loadings $e_0^\alpha$, $\alpha=1,2,3$, and assemble $\underline{\Sigma}$ from the corresponding averaged current responses using Eq.~\eqref{eq:avg_current_conductivity}.

The effective elastic response is evaluated analogously. We minimize the elastic energy with respect to a periodic displacement fluctuation $u$ under a prescribed constant macroscopic strain $\underline{\epsilon}_0$. The resulting local strain is
\[
\underline{\epsilon}
=\underline{\epsilon}_0+\tfrac12\big(\nabla u+(\nabla u)^T\big).
\]
The local stress is
\begin{equation}
\underline{\tau}
=\big(3\kappa\,\underline{I}^{\mathrm{hydro}}
+2\mu\,\underline{I}^{\mathrm{dev}}\big):\underline{\epsilon}.
\label{eq:stress}
\end{equation}
The elastic energy functional is
\begin{equation}
H_{\mathrm{elas}}[u]
=\int \underline{\tau}:\underline{\epsilon}\,d r,
\label{eq:elastic energy}
\end{equation}
where the integral is again evaluated over the periodic unit cell. The effective stiffness tensor $\underline{C}$ is defined by
\[
\langle \underline{\tau}\rangle
=\underline{C}:\underline{\epsilon}_0.
\]
Because $\underline{C}$ is represented by a symmetric $6\times6$ matrix, we solve Eq.~\eqref{eq:elastic energy} for six mutually orthogonal applied strains $\underline{\epsilon}_0^\alpha$, $\alpha=1,\ldots,6$, thereby recovering all independent components of $\underline{C}$.

The generally anisotropic unit-cell tensors $\underline{\Sigma}$ and $\underline{C}$ are converted into isotropic macroscopic properties using EMA. For an aggregate of spherical grains, the macroscopic conductivity $\sigma^\star$ satisfies the self-consistent equation
\begin{equation}
\mathrm{Tr}\!\left[
\underline{\Sigma}
(\underline{\Sigma}+2\sigma^\star\underline{I}_3)^{-1}
\right]=1.
\label{eq:EMA conductivity}
\end{equation}
For elasticity, the isotropic macroscopic stiffness is written as
\begin{equation}
\underline{C}^\star
=3\kappa^\star \underline{I}^{\mathrm{hydro}}
+2\mu^\star \underline{I}^{\mathrm{dev}}.
\label{eq:macroscopic stiffness}
\end{equation}
For brevity, we introduce the scalars
\begin{equation}
p_1=\frac{1}{3\kappa^\star+4\mu^\star},
\qquad
p_2=\frac{3(\kappa^\star+2\mu^\star)}
{5\mu^\star(3\kappa^\star+4\mu^\star)},
\label{eq:p values}
\end{equation}
and the tensor
\begin{equation}
\underline{A}
=\underline{I}^{\mathrm{sym}}
+(p_1\underline{I}^{\mathrm{hydro}}
+p_2\underline{I}^{\mathrm{dev}})
:(\underline{C}-\underline{C}^\star),
\end{equation}
where the symmetric identity tensor satisfies
\begin{equation}
    \underline{I}^{\mathrm{sym}} = \underline{I}^{\mathrm{hydro}} + \underline{I}^{\mathrm{dev}}.
    \label{eq:I_sym}
\end{equation}
Using the double contraction
\[
\underline{X}::\underline{Y}
=\sum_{ijkl}X_{ijkl}Y_{ijkl},
\]
the self-consistent EMA equations for the macroscopic elastic moduli can be written compactly as
\begin{equation}
\underline{I}^{\mathrm{hydro}}::\underline{A}^{-1}=1,
\qquad
\underline{I}^{\mathrm{dev}}::\underline{A}^{-1}=5.
\label{eq:EMA elastic moduli}
\end{equation}
In practice, $\underline{A}$ is converted into its $6\times6$ representation in the Frobenius-orthonormal basis of Eq.~\eqref{eq:3x3 basis} and inverted numerically.

The topology optimization therefore proceeds by varying the density field $\rho(\mathbf r)$ on a discretized three-dimensional grid. For each density field, we first compute the anisotropic unit-cell tensors $\underline{\Sigma}$ and $\underline{C}$ and then solve the EMA equations to obtain the corresponding isotropic macroscopic properties $\sigma^\star$, $\kappa^\star$, and $\mu^\star$. In this work, the optimization targets the trade-off between $\sigma^\star$ and $\mu^\star$, though $\kappa^\star$ may also be included in the objective if desired.

Because the density field contains one design variable at every grid point, direct optimization without gradient information would be computationally inefficient. We therefore compute the functional derivatives of the effective properties with respect to $\rho(\mathbf r)$. These derivatives, conventionally referred to as sensitivities in topology optimization, quantify how placing additional phase-1 material at each point changes the macroscopic response.

The cell-level sensitivities are obtained directly from the expressions above. Superscripts $\alpha$ denote the solution fields associated with the externally applied field associated with index $\alpha$. The external fields are chosen to be mutually orthogonal so that all independent components of the effective tensors can be recovered.

For conductivity, the sensitivity is
\begin{equation}
e_0^{\alpha}\cdot
\frac{\delta\underline{\Sigma}}{\delta\rho}
\cdot e_0^{\beta}
=
\frac{\partial j^{\alpha}}{\partial\rho}
\cdot e^{\beta},
\label{eq:sensitivity conductivity tensor}
\end{equation}
with
\[
\frac{\partial j}{\partial\rho}
=\frac{\partial\sigma}{\partial\rho}\,e.
\]
The derivative $\partial\sigma/\partial\rho$ follows directly from the chosen interpolation between the two phase conductivities. Similarly, for elasticity,
\begin{equation}
\underline{\epsilon}_0^{\alpha}
:\frac{\delta\underline{C}}{\delta\rho}
:\underline{\epsilon}_0^{\beta}
=
\frac{\partial \underline{\tau}^{\alpha}}{\partial\rho}
:\underline{\epsilon}^{\beta},
\label{eq:sensitivity stiffness tensor}
\end{equation}
where
\[
\frac{\partial\underline{\tau}}{\partial\rho}
=
\big(3\frac{\partial\kappa}{\partial\rho}
\underline{I}^{\mathrm{hydro}}
+2\frac{\partial\mu}{\partial\rho}
\underline{I}^{\mathrm{dev}}\big)
:\underline{\epsilon}.
\]
The derivatives $\partial\kappa/\partial\rho$ and $\partial\mu/\partial\rho$ likewise follow directly from the material interpolation.

The macroscopic sensitivities are then obtained by implicitly differentiating the EMA equations, Eqs.~\eqref{eq:EMA conductivity} and \eqref{eq:EMA elastic moduli}, with respect to $\rho$. For conductivity, this gives
\begin{equation}
\frac{\delta\sigma^\star}{\delta\rho}
=
\frac{
\mathrm{Tr}\!\left[
(\underline{I}_3
-\underline{\Sigma}
(\underline{\Sigma}
+2\sigma^\star\underline{I}_3)^{-1})
\frac{\delta\underline{\Sigma}}{\delta\rho}
(\underline{\Sigma}
+2\sigma^\star\underline{I}_3)^{-1}
\right]
}{
2\,\mathrm{Tr}\!\left[
\underline{\Sigma}
(\underline{\Sigma}
+2\sigma^\star\underline{I}_3)^{-2}
\right]
}.
\label{eq:sensitivity conductivity}
\end{equation}
For elasticity, we first define
\begin{equation}
\underline{U}^{x^\star}
=-\underline{A}^{-1}
:\frac{\partial\underline{A}}{\partial x^\star}
:\underline{A}^{-1},
\qquad
\underline{V}
=\underline{A}^{-1}
:(p_1\underline{I}^{\mathrm{hydro}}
+p_2\underline{I}^{\mathrm{dev}})
:\frac{\delta\underline{C}}{\delta\rho}
:\underline{A}^{-1},
\end{equation}
where $x^\star\in\{\kappa^\star,\mu^\star\}$. The elastic sensitivities can then be written compactly as
\begin{equation}
\begin{pmatrix}
\delta\kappa^\star/\delta\rho\\
\delta\mu^\star/\delta\rho
\end{pmatrix}
=
\begin{pmatrix}
\underline{I}^{\mathrm{hydro}}::\underline{U}^{\kappa^\star}
&
\underline{I}^{\mathrm{hydro}}::\underline{U}^{\mu^\star}
\\
\underline{I}^{\mathrm{dev}}::\underline{U}^{\kappa^\star}
&
\underline{I}^{\mathrm{dev}}::\underline{U}^{\mu^\star}
\end{pmatrix}^{-1}
\begin{pmatrix}
\underline{I}^{\mathrm{hydro}}::\underline{V}\\
\underline{I}^{\mathrm{dev}}::\underline{V}
\end{pmatrix}.
\label{eq:sensitivity elastic moduli}
\end{equation}

We maximize the weighted objective
\begin{equation}
L
=\lambda_\sigma\sigma^\star
+\lambda_\kappa\kappa^\star
+\lambda_\mu\mu^\star,
\label{eq:objective topopt}
\end{equation}
subject to
\[
0\le\rho\le1,
\qquad
\int\rho\,d\mathbf r=f,
\]
where $\lambda_\sigma$, $\lambda_\kappa$, and $\lambda_\mu$ control the relative importance of conductivity, bulk modulus, and shear modulus, respectively. Because the unit-cell volume is $V=1$, the integral constraint fixes the volume fraction of phase 1 to $f$. Varying the objective weights ($\lambda$'s) allows different portions of the trade-off surface between the effective properties to be explored.

We employ a density-filtered, symmetric two-phase SIMP-like penalization to optimize the relaxed density field while promoting a binary final structure. The unfiltered density $\rho(\boldsymbol{r})$ remains the design variable and is subject to the prescribed volume-fraction constraint. Before evaluating the local material properties, however, we apply a spatial density filter,
\begin{equation}
\bar{\rho}
=
\mathcal{F}_{r_{\min}}[\rho],
\end{equation}
where $\mathcal{F}_{r_{\min}}$ denotes the filtering operator with characteristic radius $r_{\min}$. The filter smooths variations in the density field, suppressing grid-scale features and introducing a minimum characteristic length scale into the optimized structures. The filtered density $\bar{\rho}\in[0,1]$ is then used to interpolate the local material properties. For each $x\in\{\sigma,\kappa,\mu\}$, we use
\begin{equation}
x(\bar{\rho})
=
x_1\bar{\rho}^{\,n_x}
+
x_2(1-\bar{\rho})^{n_x},
\qquad n_x>1,
\end{equation}
where the penalization exponent $n_x$ may depend on the property being interpolated. This expression recovers the exact phase properties at $\bar{\rho}=1$ and $\bar{\rho}=0$, while suppressing the properties of intermediate-density regions. Continuous densities therefore permit efficient gradient-based optimization, whereas the power-law penalization drives the optimized design toward a nearly binary distribution of the two constituent phases. The sensitivities with respect to $\bar{\rho}$ are propagated through the filter to obtain gradients with respect to the underlying design field $\rho$.

The sensitivities derived above provide the gradient of Eq.~\eqref{eq:objective topopt} with respect to the density at every grid point. We solve the resulting constrained problem using a projected quasi-Newton method \cite{schmidt2009}, with projection used to enforce both the pointwise bounds on $\rho$ and the prescribed volume fraction.

\section{Berryman--Milton bounds}
\label{app:BM_bounds}

We construct the Berryman--Milton (BM) comparison bounds by combining
the Beran conductivity bounds~\cite{torquato2002} with the
Milton--Phan-Thien shear-modulus bounds~\cite{milton1982},
using their shared three-point microstructural parameter
$\zeta\in[0,1]$, following~\cite{berryman88}.
Define $\overline a=f_1a_1+f_2a_2$,
$\widetilde a=f_2a_1+f_1a_2$,
$\langle a\rangle_\zeta=\zeta a_1+(1-\zeta)a_2$, and
\begin{equation}
 \mathcal F_a(q)=\overline a-
 \frac{f_1f_2(a_1-a_2)^2}{\widetilde a+q}.
 \label{eq:BM_scalar_map}
\end{equation}
The conductivity and shear bounds are parameterized by
\begin{equation}
 \begin{aligned}
 \sigma_+(\zeta)&=\mathcal F_\sigma\!\left(2\langle\sigma\rangle_\zeta\right),
 &\sigma_-(\zeta)&=\mathcal F_\sigma\!\left(\frac{2}{\langle\sigma^{-1}\rangle_\zeta}\right),\\
 \mu_+(\zeta)&=\mathcal F_\mu\!\left(T_+(\zeta)\right),
 &\mu_-(\zeta)&=\mathcal F_\mu\!\left(\frac{1}{X_-(\zeta)}\right),
 \end{aligned}
 \label{eq:BM_parametric_bounds}
\end{equation}
where
\begin{align}
 T_+(\zeta)&=\frac{24M(6B+7A)+45AB}{6(21B+2A+40M)},
 \label{eq:BM_Tplus}\\
 X_-(\zeta)&=\frac{2\{H(2D+21C)+40CD\}}{56D+48C+15H},
 \label{eq:BM_Xminus}
\end{align}
with $A=\langle\mu\rangle_\zeta$,
$B=\langle\kappa\rangle_\zeta$,
$C=\langle\mu^{-1}\rangle_\zeta$,
$D=\langle\kappa^{-1}\rangle_\zeta$,
$M=\max(\mu_1,\mu_2)$, and $H=1/\min(\mu_1,\mu_2)$.
These shear expressions follow from the Milton--Phan-Thien formulas
by maximizing the upper bound and minimizing the lower bound over
the second microstructural parameter,
$5\zeta/21\leq\eta\leq(16+5\zeta)/21$, as prescribed in the BM
construction~\citep{berryman88}.

At fixed volume fraction, we vary $\zeta$ from zero to one and plot
the four sign combinations
$(\sigma_\alpha(\zeta),\mu_\beta(\zeta))$,
$\alpha,\beta\in\{-,+\}$.
The outermost upper and lower branches give the BM cross-property bounds for the specified conductivity--elasticity ordering, intersected with the separate Hashin--Shtrikman bounds. The volume-fraction-agnostic bounds are obtained by taking the outer envelope over $f_1$.






\bibliographystyle{unsrt} 
\bibliography{refs}

\end{document}